\documentclass[fleqn,usenatbib]{mnras}

\usepackage{newtxtext,newtxmath}

\usepackage[T1]{fontenc}
\usepackage{ae,aecompl}

\pdfoutput=1 
\usepackage{amsmath,amstext}
\usepackage{longtable}
\usepackage{comment}

\usepackage{graphicx}   
\usepackage{amsmath}    
\usepackage{amssymb}    
\usepackage{pdflscape}

\title[Angular momentum with adaptive optics at $z\sim 1.5$]{Angular momentum of $z\sim 1.5$ galaxies and their local analogues with adaptive optics.}

\author[Sarah M. Sweet et al.]{
Sarah M. Sweet,$^{1,2}$\thanks{E-mail: sarah@sarahsweet.com.au} 
Deanne B. Fisher,$^{1}$
Giulia Savorgnan, 
Karl Glazebrook,$^{1}$
\newauthor 
Danail Obreschkow,$^{3}$
Steven Gillman,$^{4,5}$
Alfred L. Tiley,$^{5}$
Claudia D. P. Lagos,$^{2,3}$
\newauthor
Liang Wang,$^{3}$
A. Mark Swinbank,$^{4,5}$
Richard Bower,$^{4,5}$
Ray M. Sharples$^{6}$
\\
$^{1}$Centre for Astrophysics and Supercomputing, Swinburne University of Technology, PO Box 218, Hawthorn, VIC 3122, Australia\\
$^{2}$ARC Centre of Excellence for All Sky Astrophysics in 3 Dimensions (ASTRO 3D)\\
$^{3}$International Centre for Radio Astronomy Research, University of Western Australia, 7 Fairway, Crawley, WA 6009, Australia\\
$^{4}$Institute for Computational Cosmology, Durham University, South Road, Durham DH1 3LE, UK\\
$^{5}$Centre for Extragalactic Astronomy, Durham University, South Road, Durham DH1 3LE, UK\\
$^{6}$Centre for Advanced Instrumentation, Durham University, Department of Physics, South Road, Durham DH1 3LE, UK 
}

\date{Accepted XXX. Received YYY; in original form ZZZ}

\pubyear{2018}

\begin{document}
\label{firstpage}
\pagerange{\pageref{firstpage}--\pageref{lastpage}}
\maketitle

\begin{abstract}
We present stellar specific angular momentum $j_*$ measurements of two $z\sim 1.5$ galaxies in the KGES sample and 12 DYNAMO $z\sim 0.1$ analogues of high-redshift galaxies. We combine natural seeing integral field spectroscopic data to trace line emission out to high multiples of effective radius $r_e$, with adaptive optics assisted Keck/OSIRIS observations to trace the rapid rise in rotation curve in the inner regions. Our spaxel-wise integration method gives results that are on average within measurement uncertainty of the traditional rotation curve model method. 
At $z\sim 0$, combining GMOS and OSIRIS datasets improves the measurement uncertainty in $j_*$ from 13\% (GMOS only) or 16\% (OSIRIS only) to 10\%.  At $z\sim 1.5$, systematics allow for at best 20\% uncertainty on $j_*$.
DYNAMO analogues of high-$z$ galaxies have low $j_*$ for their stellar mass $M_*$, and low bulge-to-total light ratio $\beta$ for their $j_*/M_*$. The high-$z$ galaxy COSMOS 127977 has $j_*/M_*$ consistent with normal local disk galaxies, while UDS 78317 is consistent with local analogues. However, our high-resolution OSIRIS data reveal that UDS 78317 may be a merging system. 
We report a relationship between distance to the $\beta-j_*/M_*$ plane and the ratio of velocity dispersion to rotational velocity $\sigma/v_{max}$, where galaxies that deviate more from the plane are more dispersion-dominated due to turbulence. 
Much of the scatter in $M_*-j_*$ that is not explained by variations in the bulge-to-total ratio or evolution with redshift may be driven by increased turbulence due to star formation, or by treating mergers as rotating disks.
\end{abstract}




\section{Introduction}

A galaxy's angular momentum (AM) $J$ and mass $M$ are two of its fundamental properties, as together they trace the impact of cumulative tidal forces on that galaxy's size and density evolution \citep{Mo+1998}. Stellar AM $J_*$ {regulates disk thickness and colour} \citep{Hernandez+2006} and is a physical proxy for morphology as first shown by \citet{Fall1983} and later \citet[][hereafter RF12, OG14, C16, S18, P18, FR18]{RF12,OG14,Cortese+2016,Sweet+2018,Posti+2018,FR18}. 

It is common to remove the stellar mass scaling of $J$ and instead study stellar specific AM $j_* = J_*/M_*$. The earliest such study was conducted by \citet{Fall1983}, who found that $j_* \propto qM_*^{\alpha}$, with normalisation $q$ setting parallel tracks for early- and late-type galaxies, and slope $\alpha \approx 2/3$ in accordance with predictions for cold dark matter (CDM) haloes. RF12 later analysed the dependence of this relation on bulge-to-total light ratio $\beta$, and showed that $q$ differs between disky and bulge-dominated galaxies. Since then, 2D integral field spectroscopic (IFS) studies have confirmed the earlier findings that earlier types with larger bulges have lower $j_*$, but with important clarifications regarding the slope $\alpha$, as follows. For a subset of The HI Nearby Galaxy Survey \citep[THINGS,][]{Leroy+2008,Walter+2008} OG14 found that $\alpha \approx 2/3$ for $0 \leqslant \beta \leqslant 0.32$, but a 3D fit between $M_*$, $j_*$ and $\beta$ yields $\alpha \sim 1$ at constant $\beta$. C16, analysing galaxies observed by the Sydney-AAO Multi-object Integral field \citep[SAMI,][]{Croom+2012} Galaxy Survey \citep{Bryant+2015,Allen+2015,Sharp+2015} agreed, finding that $\alpha \gtrsim 2/3$ for single morphology classes and $\alpha \sim 1$ for late-types. More recently, S18 showed for THINGS, galaxies in RF12 and a subset of the Calar Alto Legacy Integral Field Area Survey \citep[CALIFA,][]{Sanchez+2012,Husemann+2013,Walcher+2014,Sanchez+2016} that $\alpha = 0.56 \pm 0.06$ for all bulge fractions, and $\alpha = 1.03 \pm 0.11$ for constant $\beta$ when $\beta$ is treated as a free parameter. The finding that $\alpha \sim 1$ leads to a tight relation in $\beta-j_*/M_*$ space, particularly for galaxies that host pseudobulges and have bulge-to-total mass ratios smaller than $\beta \lesssim 0.4$. Such galaxies appear to progress along this relation as they build their pseudobulges through secular evolution \citep{kormendy2004,Wyse+1997,Sweet+2018}. Conversely, FR18 did not find separate relations for galaxies that host pseudobulges and classical bulges.

At redshifts $z>1$ galaxies become increasingly disparate from traditional morphological classifications in the Hubble sequence \citep[reviewed in][]{Glazebrook2013,Madau+2014}. The dynamical time of the universe is shorter, so major and minor mergers are common \citep{Baugh+1996,Weil+1998,Tissera2000}. The first stable disks are starting to appear but have clumpy morphologies \citep{Glazebrook+1995b,Driver+1995,Abraham+1996a,Abraham+1996b,Conselice+2000,Elmegreen+2005}, high gas fractions \citep{Daddi+2010,Tacconi+2013}, high rates of star formation \citep{Bell+2005,Juneau+2005,Swinbank+2009,Genzel+2011} and corresponding enhanced turbulence with respect to local spiral galaxies \citep{ForsterSchreiber+2009,Wisnioski+2011,Wuyts+2012,Fisher+2014}. These high-$z$ galaxies are predicted in the most recent cosmological hydrodynamical simulations to have lower $j_*/M_*$ \citep[e.g.][]{Lagos+2017,Teklu+2015}, linked to lower disk stability against formation of their star-forming clumps \citep{Obreschkow+2015}. There are few analyses at high redshift, owing to the scarcity of high-resolution observations of such galaxies
, and all use the proxy $j = k r v$ for some characteristic radius $r$ and velocity $v$, with proportionality $k$ dependent on the S{\'e}rsic index \citep{Sersic1963} in an effort to account for the variation in $j_*$ with morphology
\citep{ForsterSchreiber+2006,Burkert+2016,Contini+2016,Swinbank+2017,Harrison+2017,Alcorn+2018}. 
Most find $\alpha$ consistent with 2/3, with redshift dependence varying from no evolution \citep{Burkert+2016,Alcorn+2018} up to a factor of (1+$z$)$^{-1.5}$ \citep{ForsterSchreiber+2006}.

Samples of nearby galaxies that have properties similar to those at high redshifts are easier to study than their high-$z$ counterparts, given the relative gains in surface brightness and spatial resolution \citep{Glazebrook2013}. The caveat is that local analogues may not be truly representative of high-$z$ galaxies, so it is insightful to compare the two samples where possible. Possible analogues of high-$z$ galaxies include $z\sim 0.1$ DYNAmics of Massive Objects (DYNAMO) sample \citep{Green+2010,Green+2014} - the turbulent disk galaxies in DYNAMO are analogous to the clumpy disks with high star-formation rates found at $1\lesssim z\lesssim 2$;
Lyman-break analogues (LBAs), which are similar to Lyman-break galaxies at $z\sim 3$ in terms of UV luminosity, stellar mass and star-formation rate \citep{Heckman+2005}; 
supercompact LBAs \citep{Basu-Zych+2009,Goncalves+2010}, with high velocity dispersions similar to the $z\sim 2$ galaxies of \citet{Law+2007};
tadpole galaxies \citep{Straughn+2006,Elmegreen+2010}, which each have a single bright star-forming clump with a tail, and appear to be smaller versions of tadpole galaxies at high redshift as first identified by \citet{vandenBergh+1996};
green peas \citep{Cardamone+2009}, which are compact and low-mass but have high star-formation rates and velocity dispersions, and clumpy morphology. Of these, only DYNAMO has extensive IFS data including an analysis of AM, and then only for four galaxies \citep{Obreschkow+2015}.

Whether locally or at high redshift, $j_*$ is difficult to measure. The best practise is to integrate over spatially-resolved $J_i$ in spaxels $i$ from resolved velocity and mass maps, as in OG14, C16 and S18. OG14 demonstrated that using 2D IFS affords an order-of-magnitude improvement in precision over integrated or long-slit spectroscopic observations of local galaxies. Ideally, one aims to reach large multiples of the effective radius $r_e$ in order to trace the bulk of $j_*$, e.g. 0.99$j_*$ is enclosed within 3$r_e$ \citep{Sweet+2018}. To do so requires sufficient signal-to-noise at the faint outskirts of the galaxy. One also desires to adequately sample the inner regions of the galaxies where the rotation curve is rapidly rising, in order to sufficiently constrain the velocity field. This requires adaptive optics (AO)-assisted IFS observations, but the improved PSF and finer sampling come at the price of signal-to-noise, so such data are less suitable for probing to large multiples of $r_e$, as discussed in \citet{Glazebrook2013}. (One exception may be the recent deep AO imaging for the SINS/zC-SINF survey at $z\sim 2$ \citep{ForsterSchreiber+2018}, but AM measurements have not yet been presented.) The combination of seeing-limited and AO-assisted IFS data to measure $j_*$ was first demonstrated by \citet[][hereafter O15]{Obreschkow+2015} for four galaxies in DYNAMO. O15 found that their $j_*$ is three times lower for their $M_*$ than normal local galaxies. These analogues have $\beta < 0.1$ so their low $j_*$ is not a consequence of their photometric morphology but may be related to their star-formation-induced turbulence. 

In this work we combine natural seeing data to trace low surface brightness outskirts of the galaxies and the bulk of $j_*$, with adaptive optics (AO)-assisted data to mitigate the effects of beam-smearing in the high surface brightness inner regions, giving improved constraints on the velocity field. We present the first such measurement for galaxies at high redshift ($z\sim 1.5$) along with measurements for 12 local analogues from the DYNAMO sample. 

In Section~\ref{sec:sample} we describe the samples and our datasets. Section~\ref{sec:methods} contains the details of our methods for making the measurements presented in this paper. In Section~\ref{sec:dataset} we analyse the relative merits of seeing-limited and AO-assisted data, and the combination of the two. We analyse the relation between stellar mass, specific AM and morphology for high-$z$ galaxies and their local analogues in Section~\ref{sec:bt_jm}. Section~\ref{sec:discussion} discusses possible future evolution of DYNAMO galaxies and the implications in light of our $z\sim 1.5$ observations. Section~\ref{sec:conclusion} concludes the paper.

We assume a cosmology where $H_0 = 70 \rm{km s}^{-1} \rm{Mpc}^{-1}$, $\Omega_M = 0.27$ and $\Omega_\Lambda = 0.73$, and quote comoving coordiates.

\section{Sample \& Observations}
\label{sec:sample}

We compare the angular momentum properties of two galaxies at $z\sim 1.5$ with 12 local turbulent galaxies in DYNAMO, which have been suggested as local analogues of high-$z$ galaxies, and normal local galaxies from THINGS, RF12 and CALIFA which were presented in S18. The sample selection, observations and data processing for the DYNAMO and $z\sim 1.5$ samples are described in this Section.

\subsection{DYNAMO -- low-redshift analogue sample}
The DYNAMO sample \citep{Green+2010,Green+2014} is a set of 95 star-forming galaxies at $z \sim$ 0.1 selected from the Sloan Digital Sky Survey Data Release 4 \citep{Adelman-McCarthy+2006} as having high H$\alpha$ fluxes due to star formation $L_{H\alpha} > 10^{42}\ \rm{erg}\ \rm{s}^{-1}$. DYNAMO galaxies are analogous to rotating disk galaxies at $1 \lesssim z \lesssim 2$ in that they have similarly high velocity dispersions \citep{Green+2010,Bassett+2014} and clumpy morphologies \citep{Fisher+2017b}. Their specific star formation rates match those of galaxies between $0 \lesssim z \lesssim 2$ \citep{Green+2010,Green+2014}, Fisher et al., in prep.).

For this and related projects we observed a subset of 20 DYNAMO galaxies. Seeing-limited observations were obtained with Gemini GMOS \citep{Hook+2004} for 13 of the 20, and emission line intensity and velocity maps measured at H$\beta$ $\lambda$~4861~\AA\ \citep{Fisher+2017}. Keck OSIRIS \citep{Larkin+2006,Larkin+2006SPIE} AO observations covering the P$\alpha$ $\lambda$~18750~\AA\ line were obtained for another 13 of the sample \citep{Oliva-Altamirano+2018}; seven have both GMOS and OSIRIS data. 

\citet{Fisher+2017b} presented Hubble Space Telescope (HST) narrow-band H$\alpha$ and continuum imaging for 10 of the galaxies, using the FR647M, FR716N and FR782N ramp filters on the Wide Field Camera / Advanced Camera for Surveys (ACS). 
This imaging is used to constrain inclination and scale length, and for surface density images, {for which we assume a constant mass-to-light ratio. \footnote{As we are concerned with \emph{specific} angular momentum, the mass normalisation cancels in our calculations.} We use medium-band imaging for the galaxies that have those data, and narrow-band otherwise.} OSIRIS 1.9$\mu$m maps are used for the galaxies that do not have any HST imaging.

\subsection{KGES -- high-redshift sample}
Our high-redshift sample is drawn from the KMOS Galaxy Evolution Survey (KGES; Tiley et al., in preparation). KGES comprises KMOS \citep[][]{Sharples+2013} observations of H$\alpha$, [N {\sc ii}]6548 and [N {\sc ii}]6583 emission from 285 galaxies at $1.3 \lesssim z \lesssim 1.5$ in well-known extragalactic fields (COSMOS, CDFS, and UDS). Target galaxies were predominantly selected to be bright ($K>22.7$) and blue ($I-J<1.7$), with higher priority assigned to those which have an established spectroscopic redshift. The KMOS PSF FWHM for the targets in this work is 0".6.

We observed two KGES targets, COSMOS 127977 and UDS 78317, with Keck OSIRIS during 2017 December 5-7 in the Hn4 and Hn3 filters respectively, in order to cover rest-frame H$\alpha$ $\lambda$~6563~\AA. These data were processed using the current OSIRIS data reduction pipeline DRP 4.0.0 using rectification matrices taken on 2017 December 14-15. Emission line intensity and velocity maps were then extracted from the data cubes. The OSIRIS PSF FWHM for these observations is 0".1.

{Surface density maps are derived from} HST ACS $I$-band F814W archival imaging {assuming a constant mass-to-light ratio}, with COSMOS 127977 imaging from the Cosmic Evolution Survey \citep[COSMOS,][]{Scoville+2007} (HST program 9822), and UDS 78317 imaging from the Cosmic Assembly Near-IR Deep Extragalactic Legacy Survey \citep[CANDELS,][]{Koekemoer+2011} (HST program 12064).

\begin{figure*}
    \includegraphics[width=0.3\linewidth]{COSMOS_127977_panel_0}
    \includegraphics[width=0.3\linewidth]{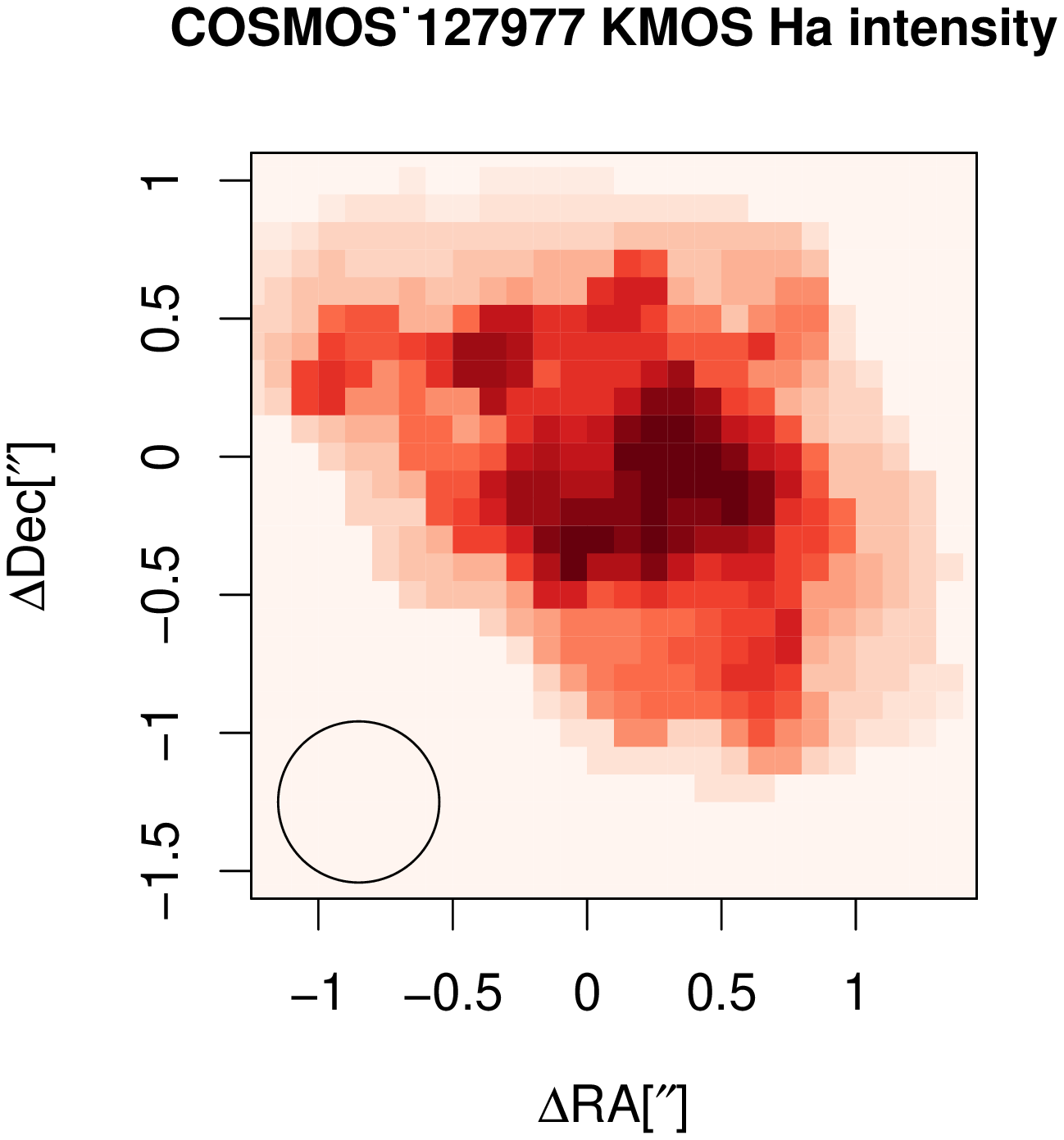}
    \includegraphics[width=0.328\linewidth]{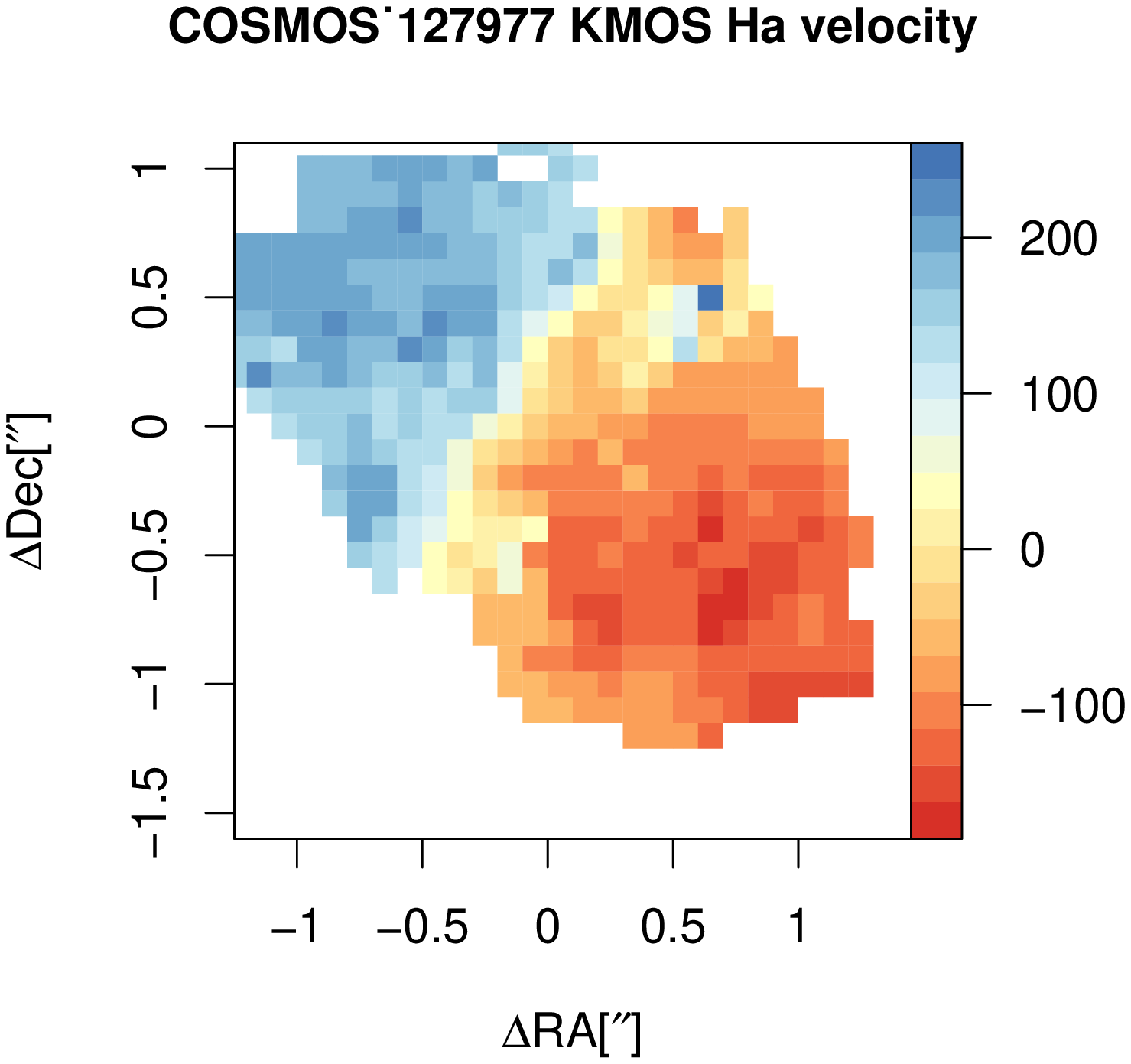}
    \includegraphics[width=0.3\linewidth]{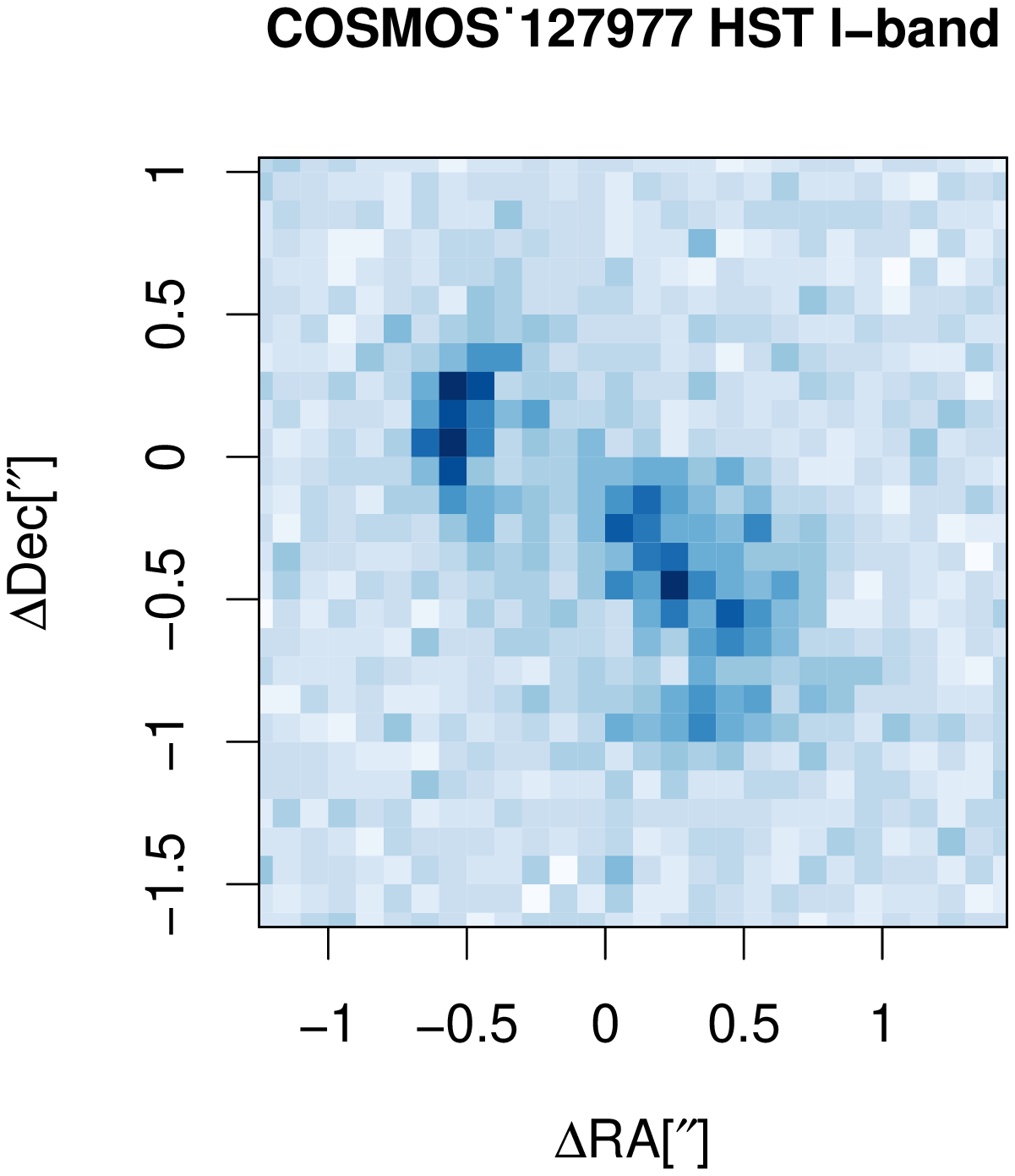}
    \includegraphics[width=0.3\linewidth]{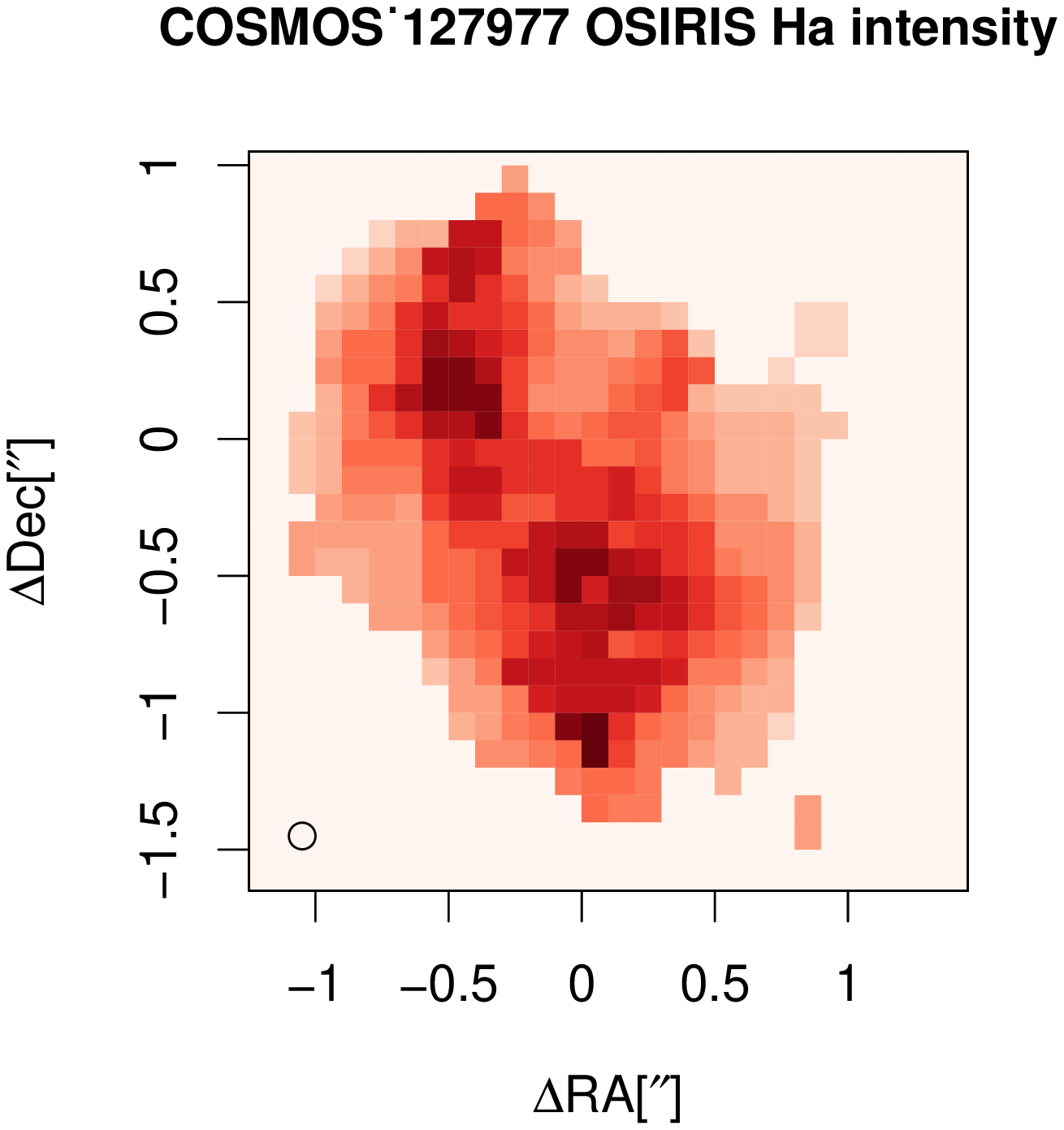}
    \includegraphics[width=0.328\linewidth]{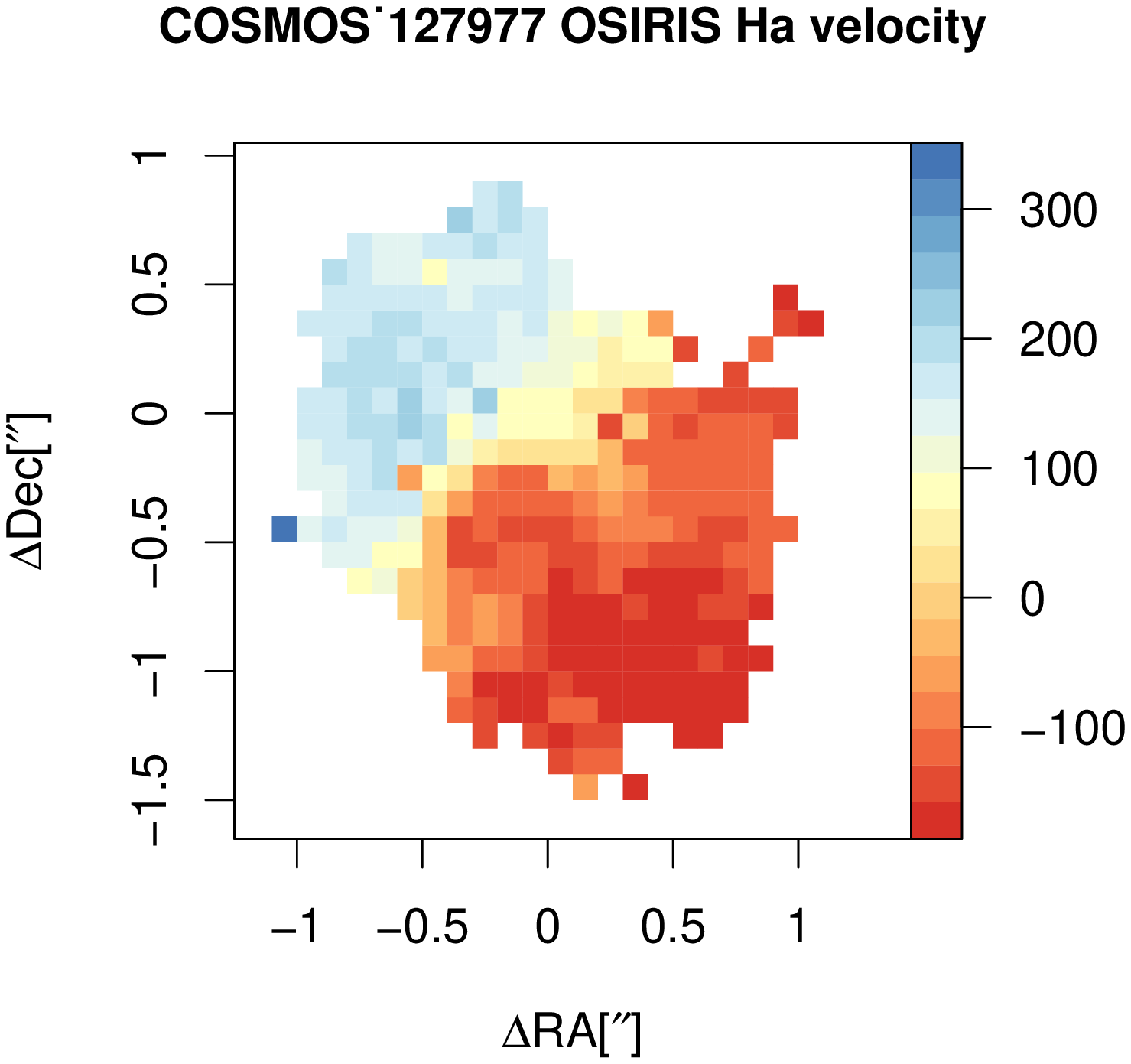}
    \caption{Ionized gas and continuum maps of COSMOS 127977 at $z = 1.62$.
    Top row: KMOS H$\alpha$ intensity, H$\alpha$ velocity. Bottom row: HST $I$-band continuum, downsampled to 0.1x0.1" spaxels, OSIRIS H$\alpha$ intensity, OSIRIS H$\alpha$ velocity. The circle in the centre panels denotes PSF FWHM. The natural seeing maps probe higher radii, while the adaptive optics-assisted maps detect more structure due to the finer PSF.}
    \label{fig:cosmos_maps}
\end{figure*}

\begin{figure*}
    \includegraphics[width=0.3\linewidth]{UDS_78317_panel_0}
    \includegraphics[width=0.3\linewidth]{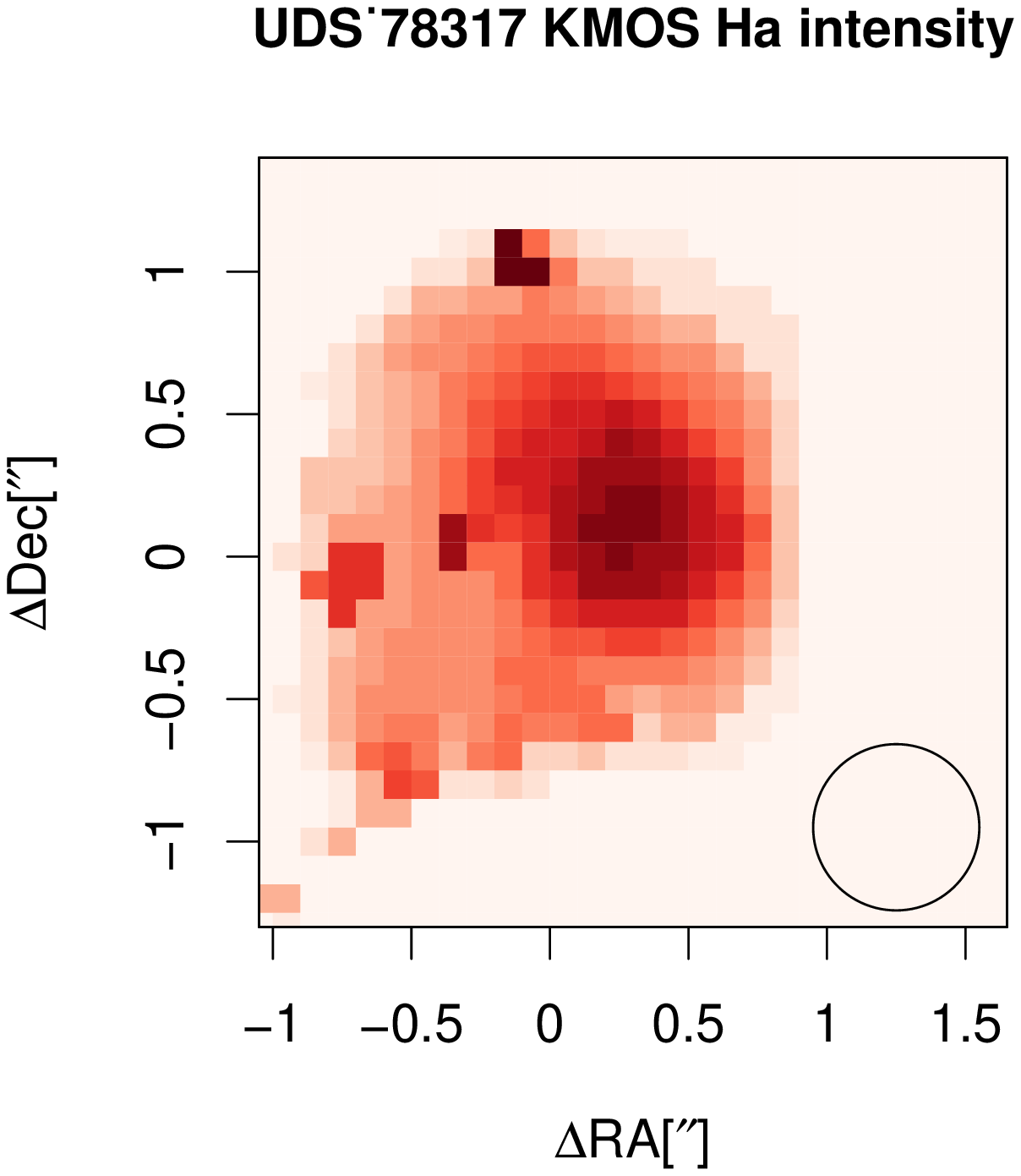}
    \includegraphics[width=0.328\linewidth]{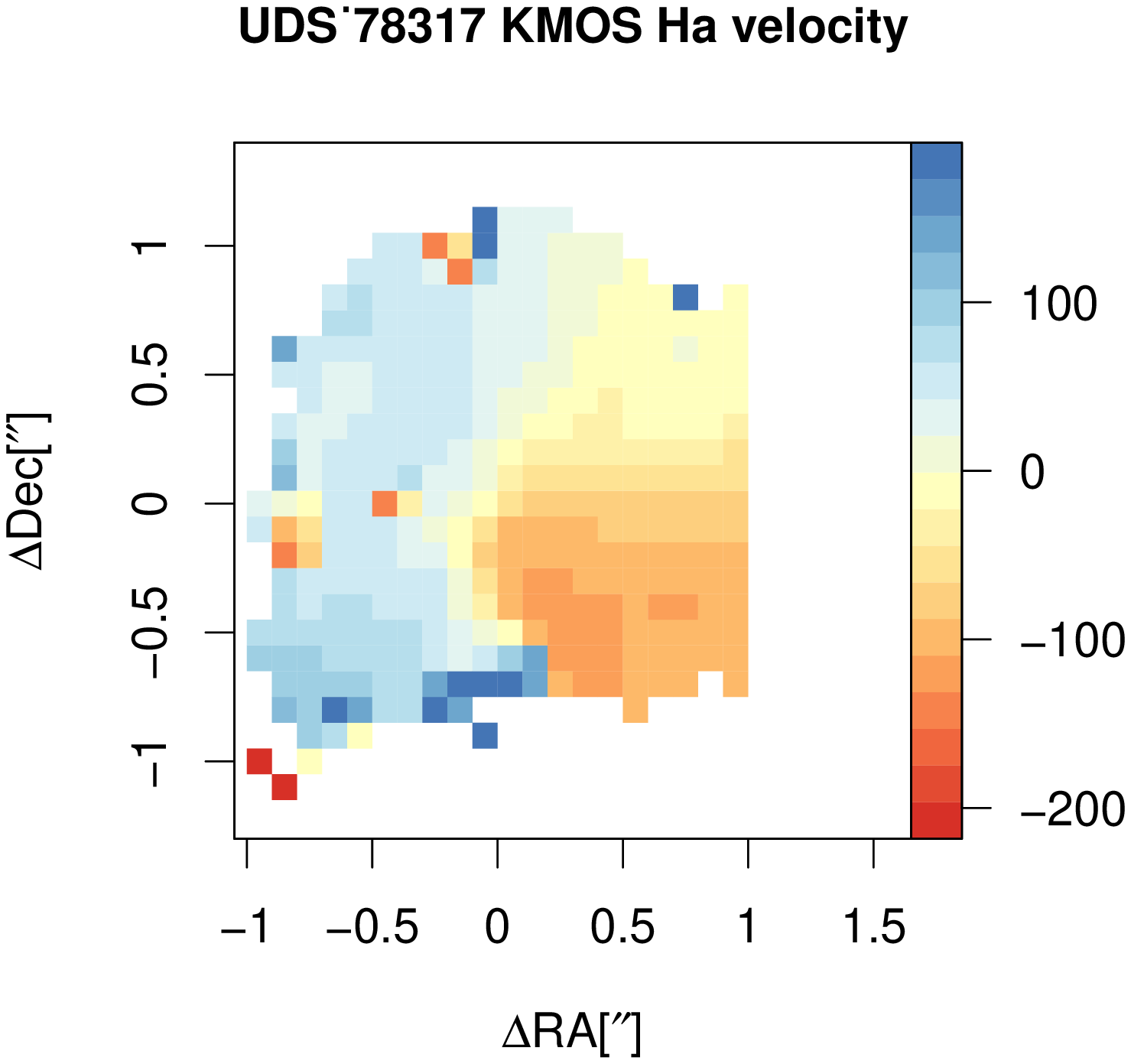}
    \includegraphics[width=0.3\linewidth]{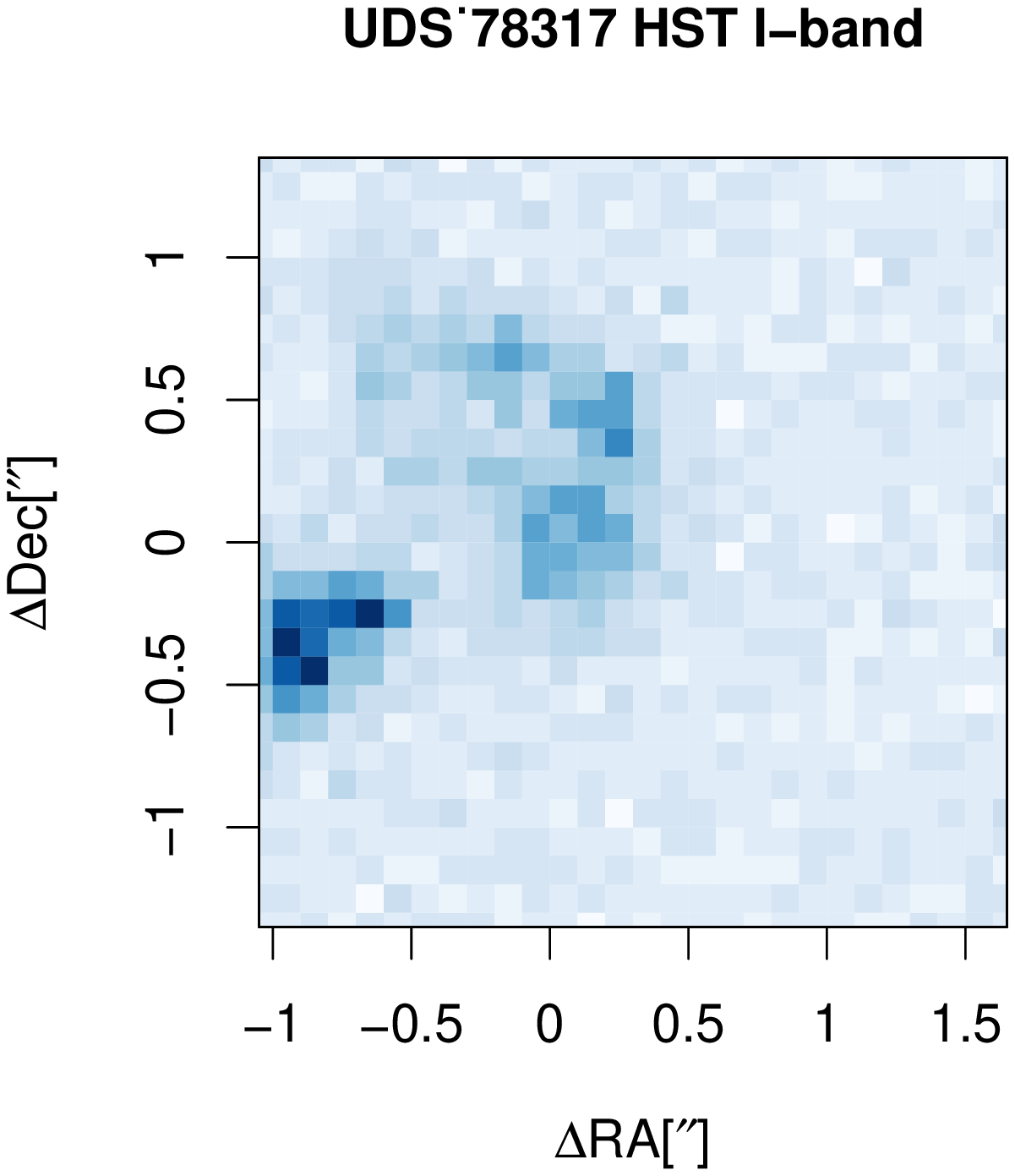}
    \includegraphics[width=0.3\linewidth]{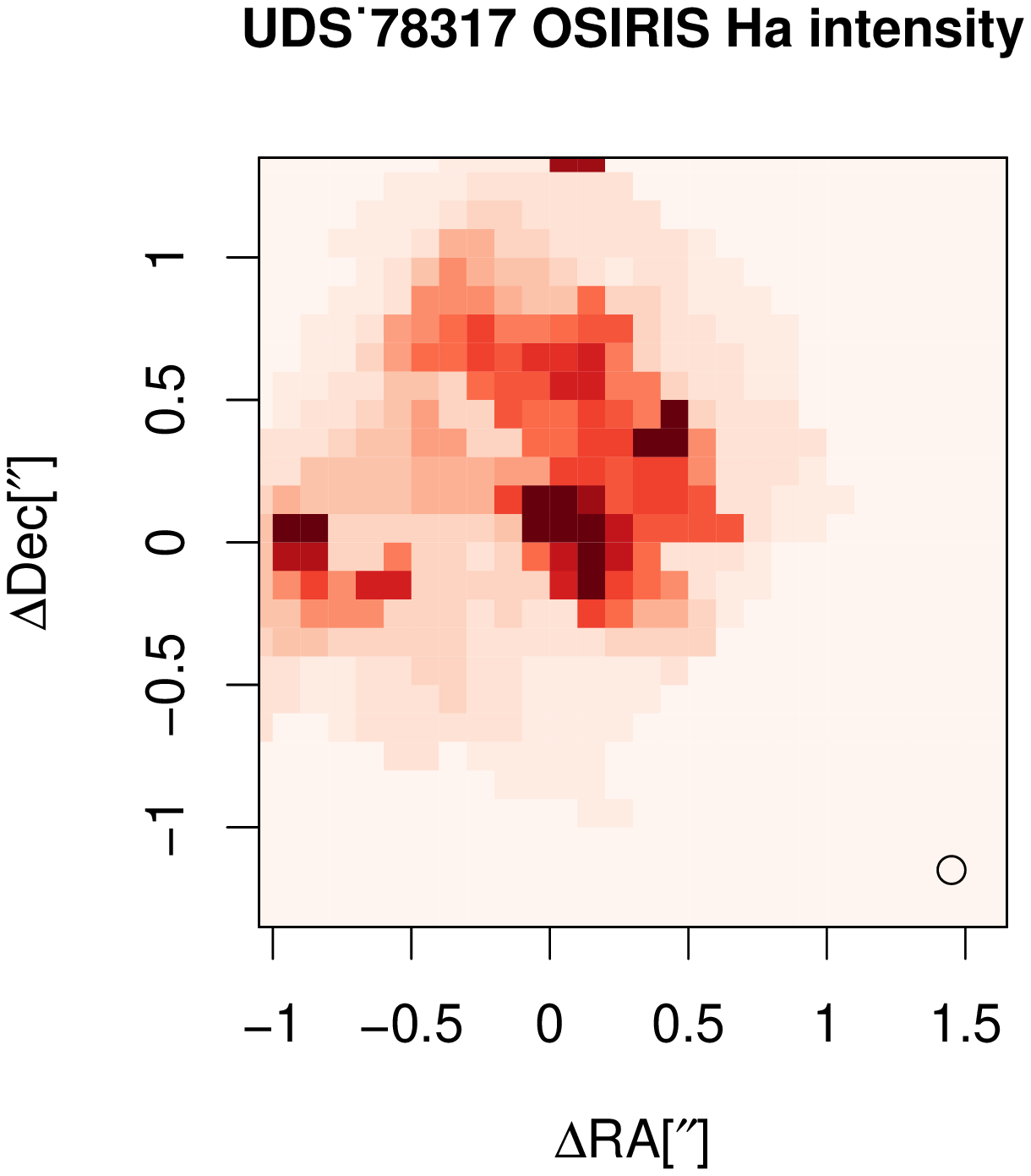}
    \includegraphics[width=0.328\linewidth]{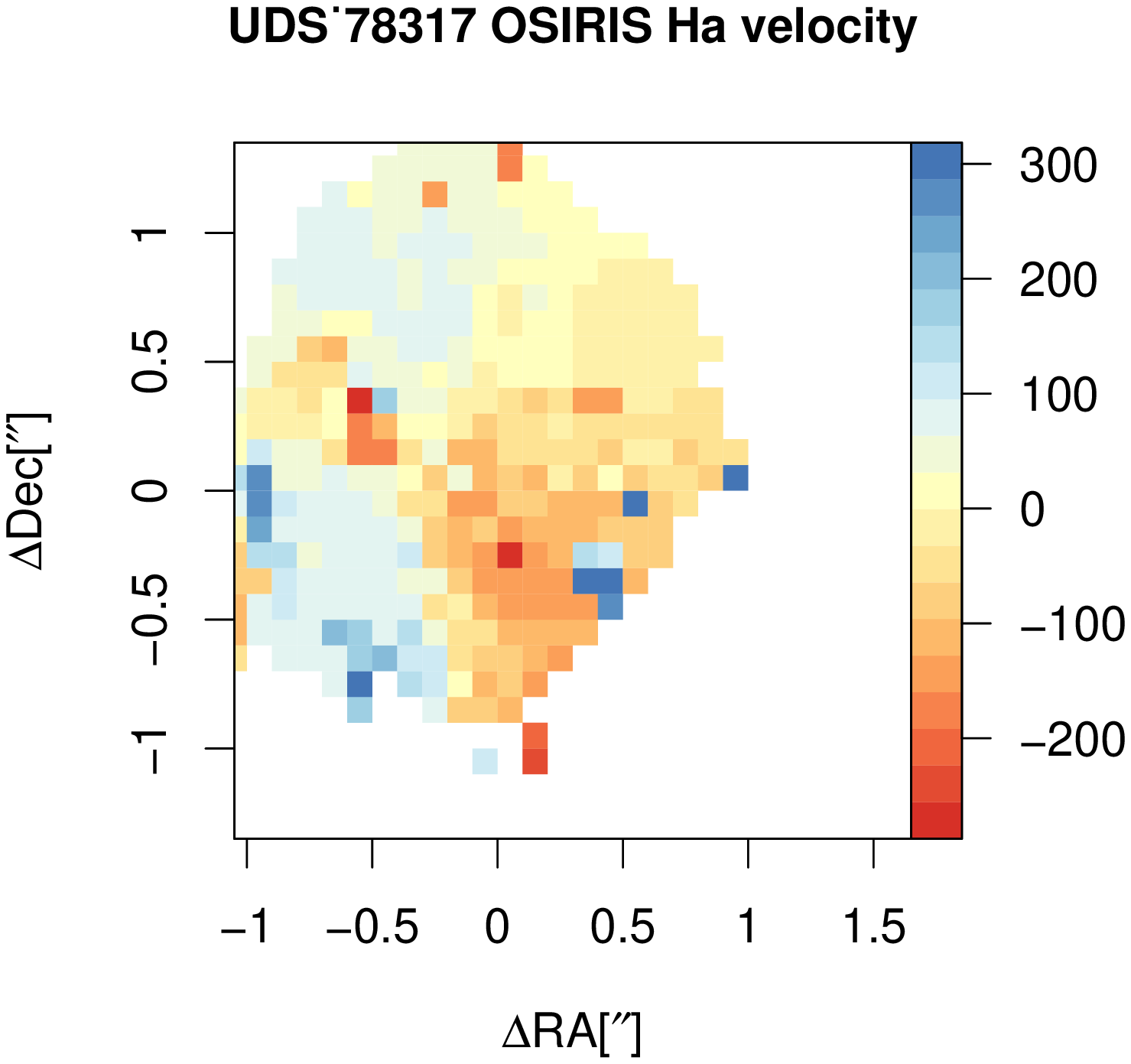}
    \caption{As for Fig.~\ref{fig:cosmos_maps} but for UDS 78317 at $z = 1.47$.}
    \label{fig:uds_maps}
\end{figure*}

Maps of COSMOS 127977 and UDS 78317 are shown in Fig.~\ref{fig:cosmos_maps} and~\ref{fig:uds_maps}, illustrating that the seeing-limited data probe to higher radii, while the AO-assisted data are more sensitive to structure, particularly in the inner regions of the galaxy. 

\section{Methods}
\label{sec:methods}

\subsection{Specific angular momentum}

For each galaxy we combine two datasets to calculate $j_*$ from spatially-resolved integral field spectroscopic (IFS) observations, using the methods of OG14, O15 and S18. 
The two complementary datasets are a) adaptive optics (AO)-assisted OSIRIS observations, which are sensitive to the rapidly-changing inner regions of the rotation curve, minimising the effects of beam smearing; b) seeing-limited GMOS/KMOS data, which are more sensitive than OSIRIS to the low surface brightness outer regions where the rotation curve becomes flat, in an effort to trace the bulk of the AM.  In the spaxels where both datasets have S/N$<$3 we include a model estimate to extrapolate the surface brightness and velocity profiles to ${\bf r}_i = \infty$. 

Below we describe the steps taken in our calculation:

1)  The observed deprojected spaxel-wise AM $J_i$ is derived separately for the AO-assisted and the seeing-limited kinematic data, where the calculation $J_i = {\bf r}_i {\bf v}_i m_i$ is performed in every spaxel $i$ at deprojected radius ${\bf r}$ whose circular velocity ${\bf v}$ is derived from ionized gas kinematic maps\footnote{\label{foot:corotation} We assume that the ionized gas corotates with the stars, expecting that the asymmetric drift between the two is negligible due to their comparable velocity dispersions \citep{Bassett+2014}. At high redshift the validity of this assumption is an open question. \citet{El-Badry+2018} found that the assumption of corotation tends to cause $j_*$ to be overestimated by around 20 per cent for galaxy disk components. If the assumption of cororation is invalid for our bulgeless $z\sim 1.5$ sample galaxies to a similar extent as it is for FIRE galaxies, then $j_*$ would similarly be reduced by 0.1 dex and the main conclusions of the paper would be unchanged.} and mass $m$ from stellar surface density maps derived from the HST images described in the previous Section, assuming a constant mass-to-light ratio. 
The kinematic centre is computed by minimising the convolution of the light-weighted velocity field with its 180-degree rotation. Inclination and position angle are assumed to be constant with radius and are derived from a fit to the HST imaging. We do not treat non-circular motions in this paper; in a future paper we will investigate the contribution of non-circular motions to total $j_*$ and spatially-resolved PDF($j_*$).

\begin{figure*}
\includegraphics[width=0.6\linewidth]{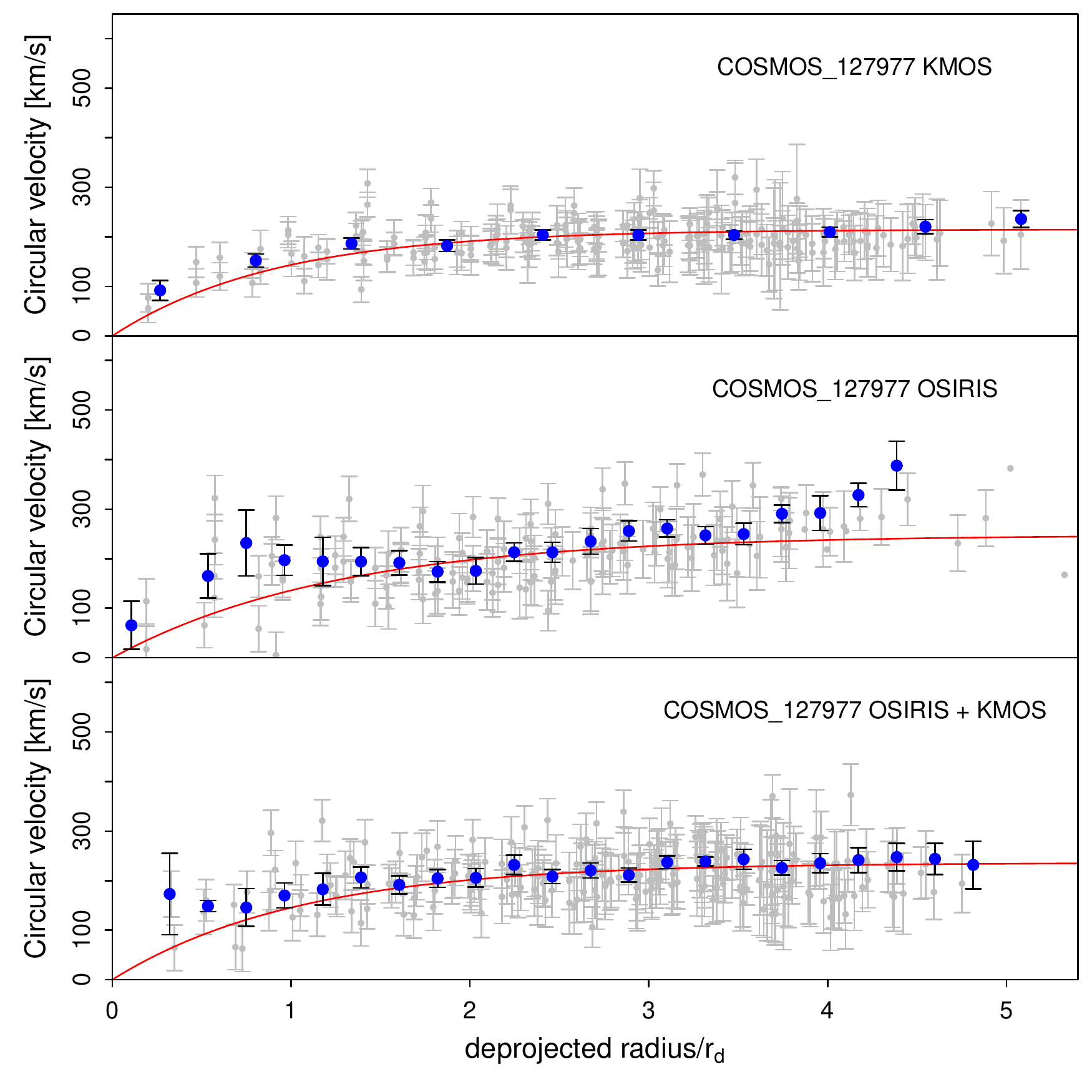}
\includegraphics[width=0.6\linewidth]{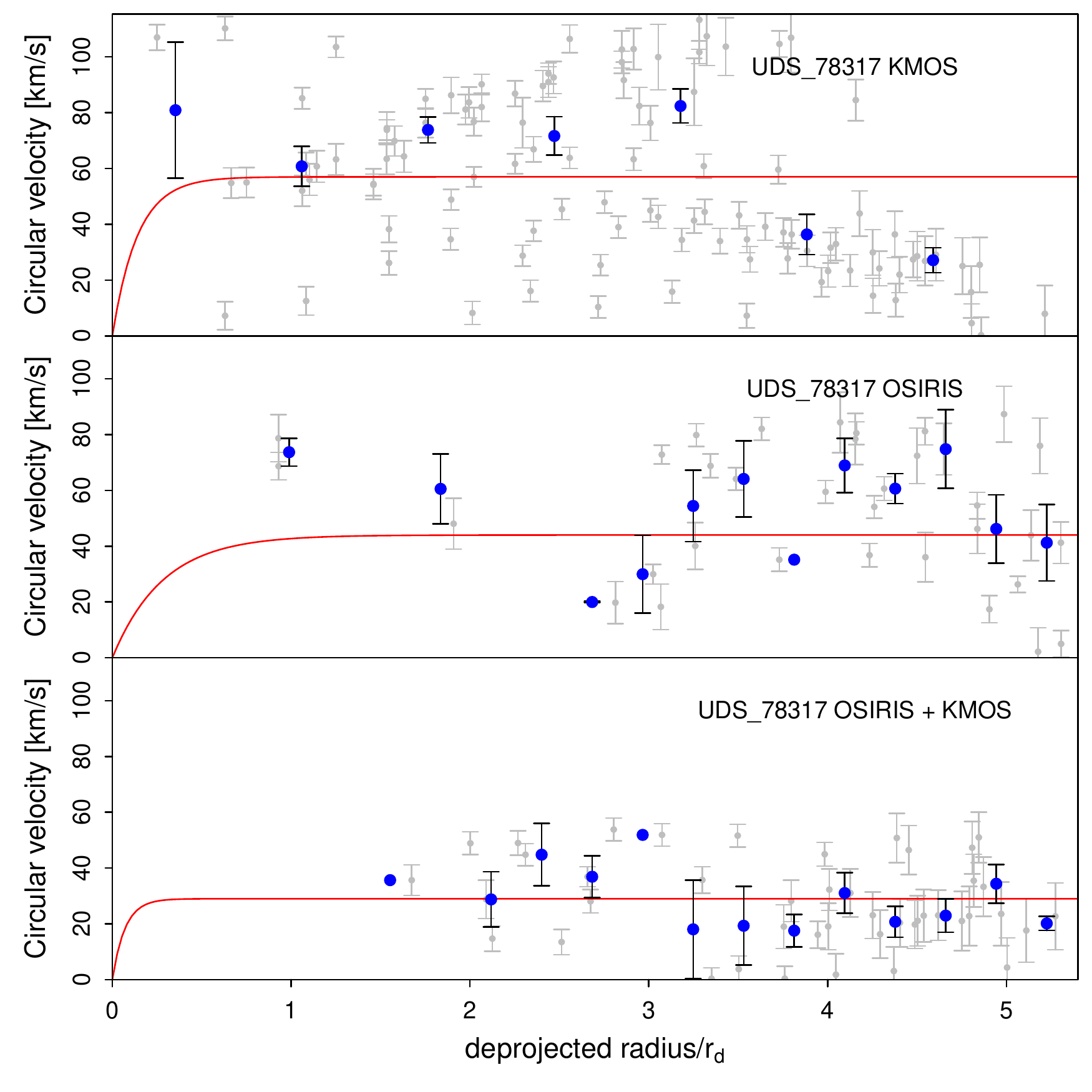}
    \caption{Rotation curves for natural seeing (top panel), adaptive optics (AO)-assisted (middle panel), and natural seeing \& AO combined data (bottom panel). Grey points are the unbinned spaxels {(after a sigma-clip for clarity only)}, blue points are binned spaxels, red curve is the model fit used to extrapolate when calculating total $j_*$. Top: COSMOS 127977. The natural seeing data are more sensitive to the low surface brightness outer regions of the galaxies, while the AO-assisted data are less affected by beam smearing, particularly in the inner parts where the rotation curves are rapidly rising. The combination of the two datasets allows for better characterisation of total stellar specific angular momentum $j_{tot}$. Bottom: UDS 78317. This galaxy is not well fit by the model rotation curve since it is not a regular disk but a merger, as revealed by the AO maps shown in Figure~\ref{fig:uds_maps}.
}
    \label{fig:rot_curves_cos}
%
\end{figure*}

2) The model ${J}_i$ in every spaxel is also computed by fitting an exponential profile ${\bf \tilde{v}}_i \approx v_{flat} \left(1 - \rm{exp}(-{\bf r}_i/r_{flat})\right)$ to the velocity field, where $r_{flat}$ is the radius at which the velocity reaches the converged velocity $v_{flat}$. 
The surface mass density is estimated by fitting $\tilde{\Sigma}({\bf r}_i) \approx s_d \rm{exp}(-{\bf r}_i/r_{d})$ to the imaging, where $s_d$ is the fitted surface mass density normalisation and $r_d$ is the exponential disk scale length. The resulting model is then 
\begin{equation}
\tilde{J}_i = {\bf r}_i {\bf \tilde{v}}_i \tilde{\Sigma}({\bf r}_i) = {\bf r}_i v_{flat} \left(1 - \rm{exp}(-{\bf r}_i/r_{flat})\right) s_d \rm{exp}(-{\bf r}_i/r_{d}) .
\end{equation}
The fitted rotation curves used to derive the model are shown in Figure~\ref{fig:rot_curves_cos}. On average, the model is consistent with the observed $J_*$ to the 5\% level, when integrating over the same high signal-to-noise spaxels. We reiterate that the model simply serves as an estimate of $J_i$ in the low signal-to-noise spaxels and allows to reach the total AM. 

3) The total $j_*$ is then given by combining 1) and 2) to calculate $J_*/M_*$, where $J_* = \lvert\sum^i {J}_i\rvert$ is the norm of the sum over:

a) the observed $J_i$ from AO-assisted data in the spaxels where the AO-assisted data have S/N$>3$,

b) the observed $J_i$ from seeing-limited data in the spaxels where the seeing-limited data have S/N$>3$ and AO S/N$<3$, and,

c) the estimated $J_i$ in other spaxels, integrated to ${\bf r}_i=\infty$.

In this way, the AO data contribute in the inner regions, the seeing-limited data contribute in the outer regions where the AO data are missing or lack the sensitivity to be reliable, and the model contributes elsewhere. The natural seeing ${\bf v}_i$ measurements are only used in the outer regions where $d{\bf v}/d{\bf r}$ is small, so the effect of beam-smearing is also small.
Including the estimated $J_i$ in the spaxels where data are missing comprises an average of 13 per cent of the total $j_*$. In Figure~\ref{fig:nat_ao} and~\ref{fig:nat_ao2} we show the cumulative stellar specific AM as a function of radius in order to illustrate the contribution to $j_*$ by the AO-assisted data, the natural seeing data and the model-informed estimate. \\

\begin{figure*}
\includegraphics[width=\columnwidth]{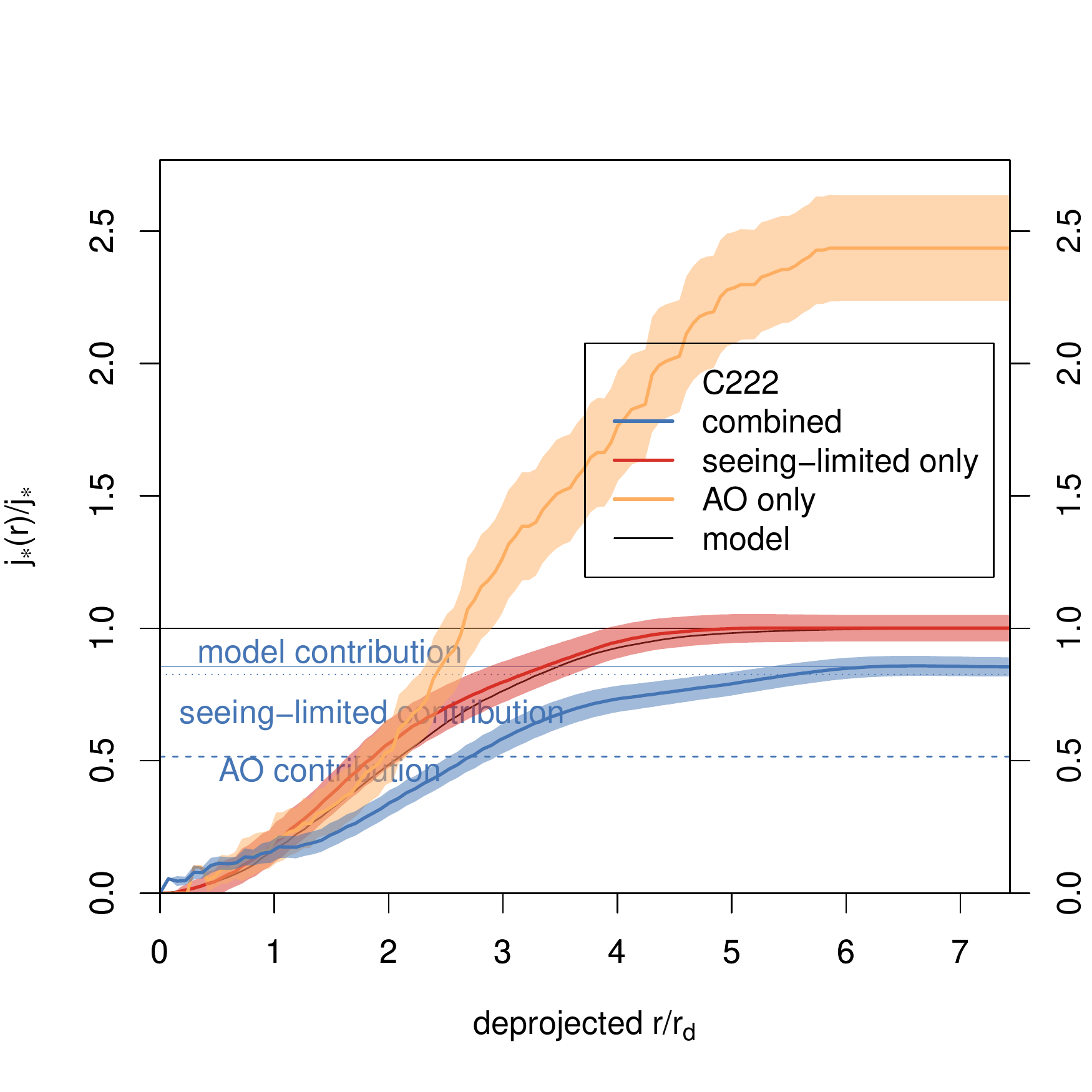}
\includegraphics[width=\columnwidth]{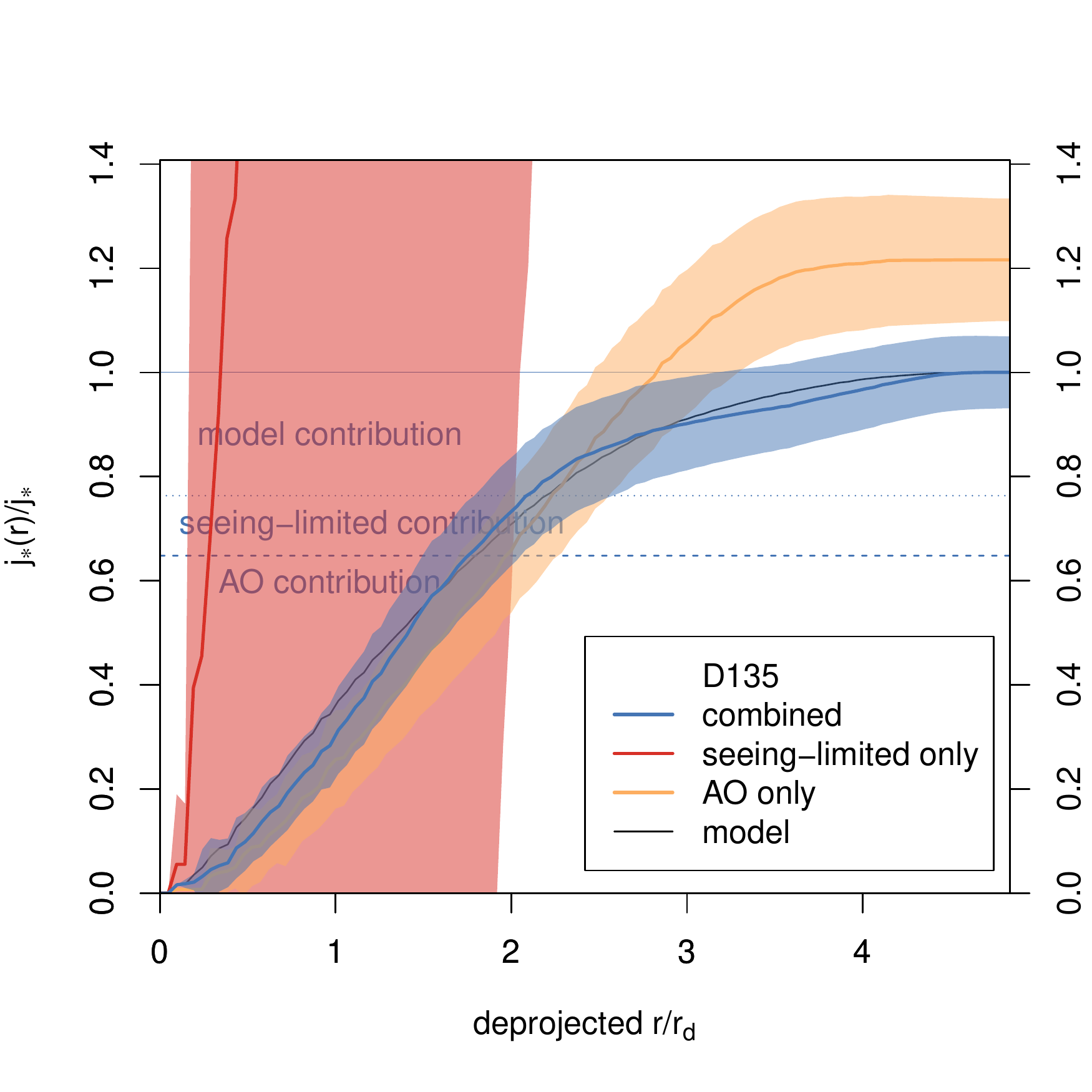}
\includegraphics[width=\columnwidth]{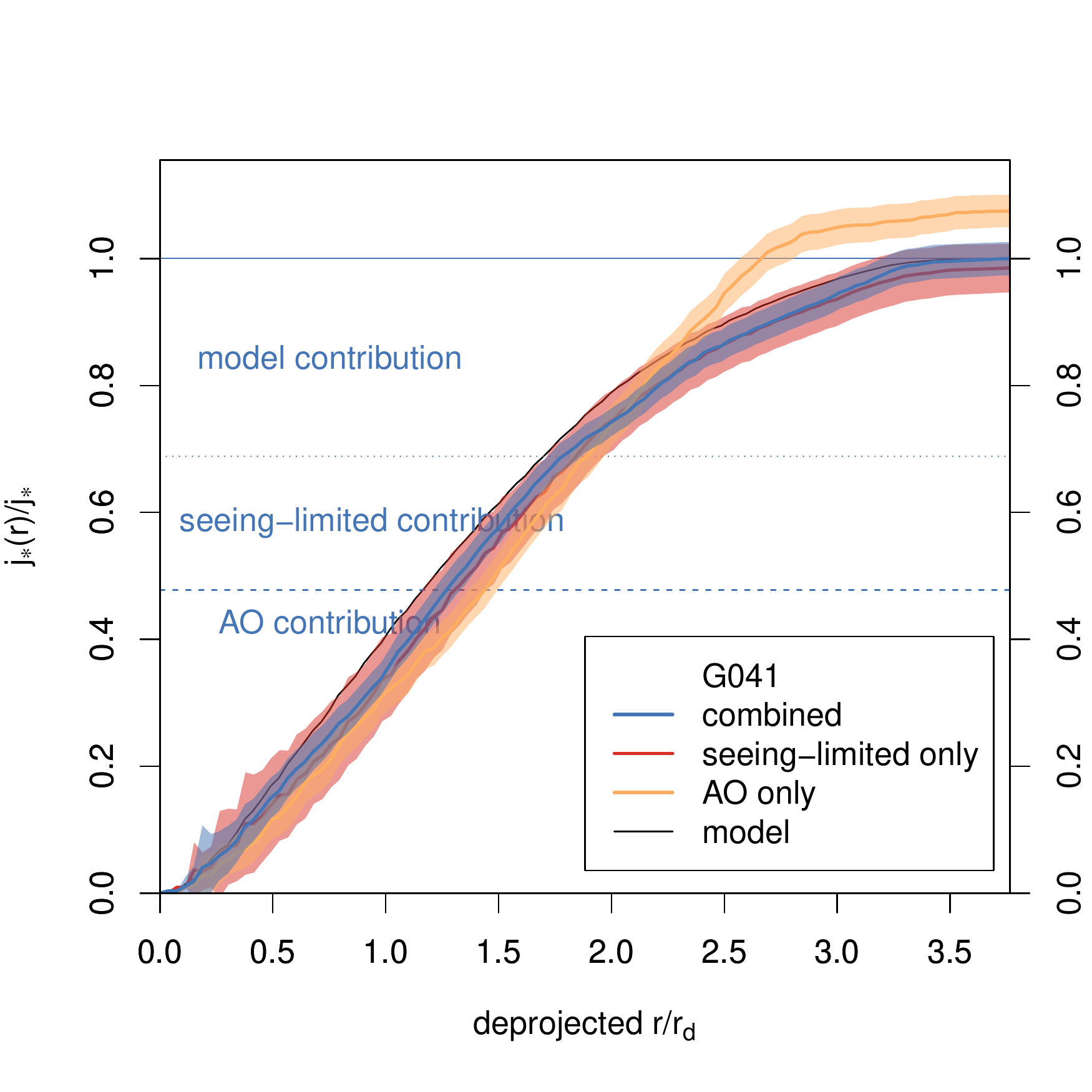}
\includegraphics[width=\columnwidth]{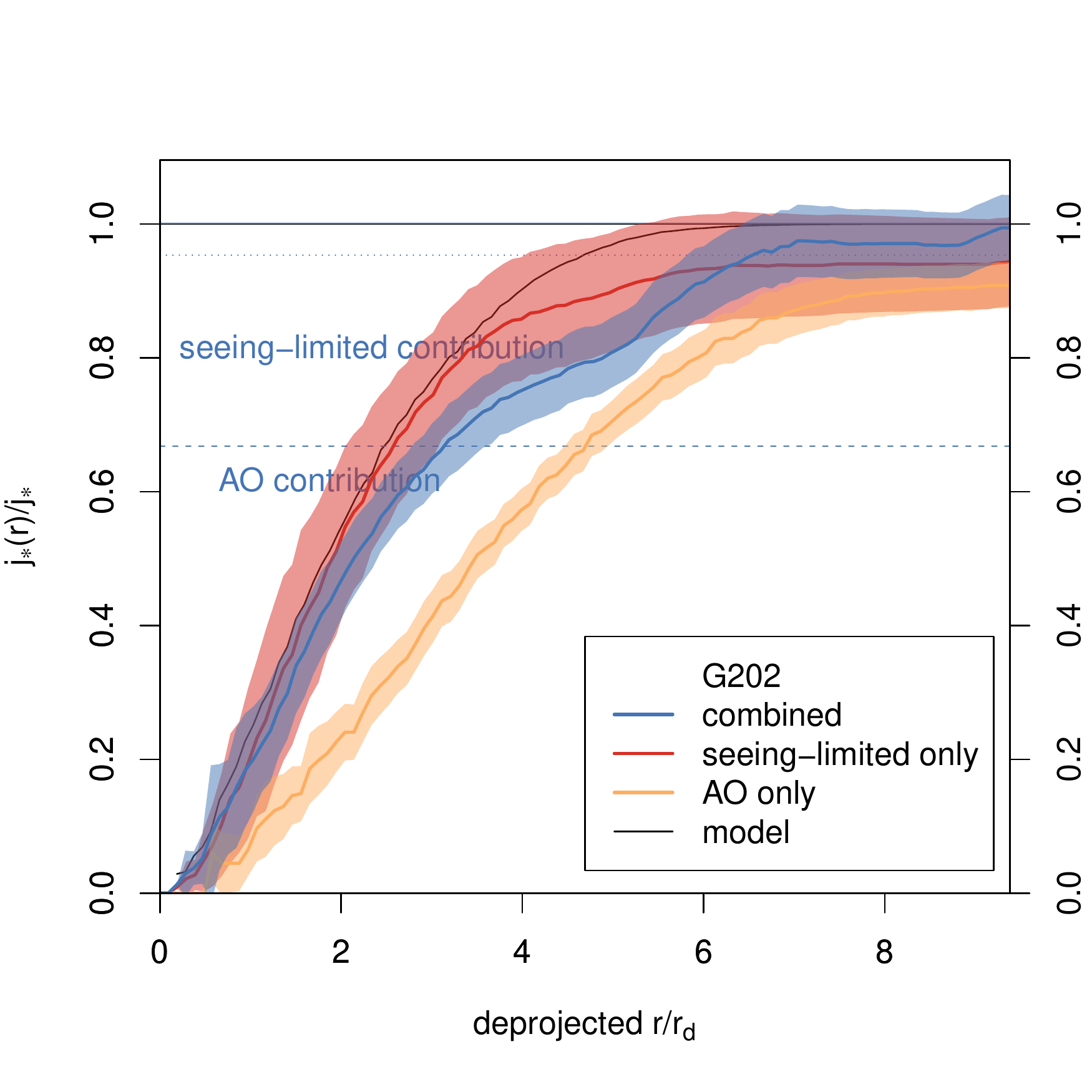}
\caption{Cumulative specific angular momentum as a function of radius. {The blue profile represents the combined dataset, with labels} showing the contribution from adaptive optics-assisted data, seeing-limited data and model-informed estimate. {The orange and red profiles indicate the same measurement made solely with seeing-limited or AO-assisted data respectively. The transparent shading indicates the uncertainty on each cumulative profile.} {The black line gives the model profile. The y-axis is normalised to the adopted $j_*$ measurement as given in Table~\ref{tab:adopted}. The horizontal lines represent the relative contribution of each component to the combined (blue) profile; due to our spaxel-wise integration method, which accounts for azimuthal variation in S/N of the various datasets, these are simply an approximate mean boundary rather than a strict radial cut.}
Only the galaxies with both natural seeing and AO data are shown in this Figure. Section~\ref{sec:dataset} has more details.}
\label{fig:nat_ao}
\end{figure*}

\begin{figure*}
\includegraphics[width=\columnwidth]{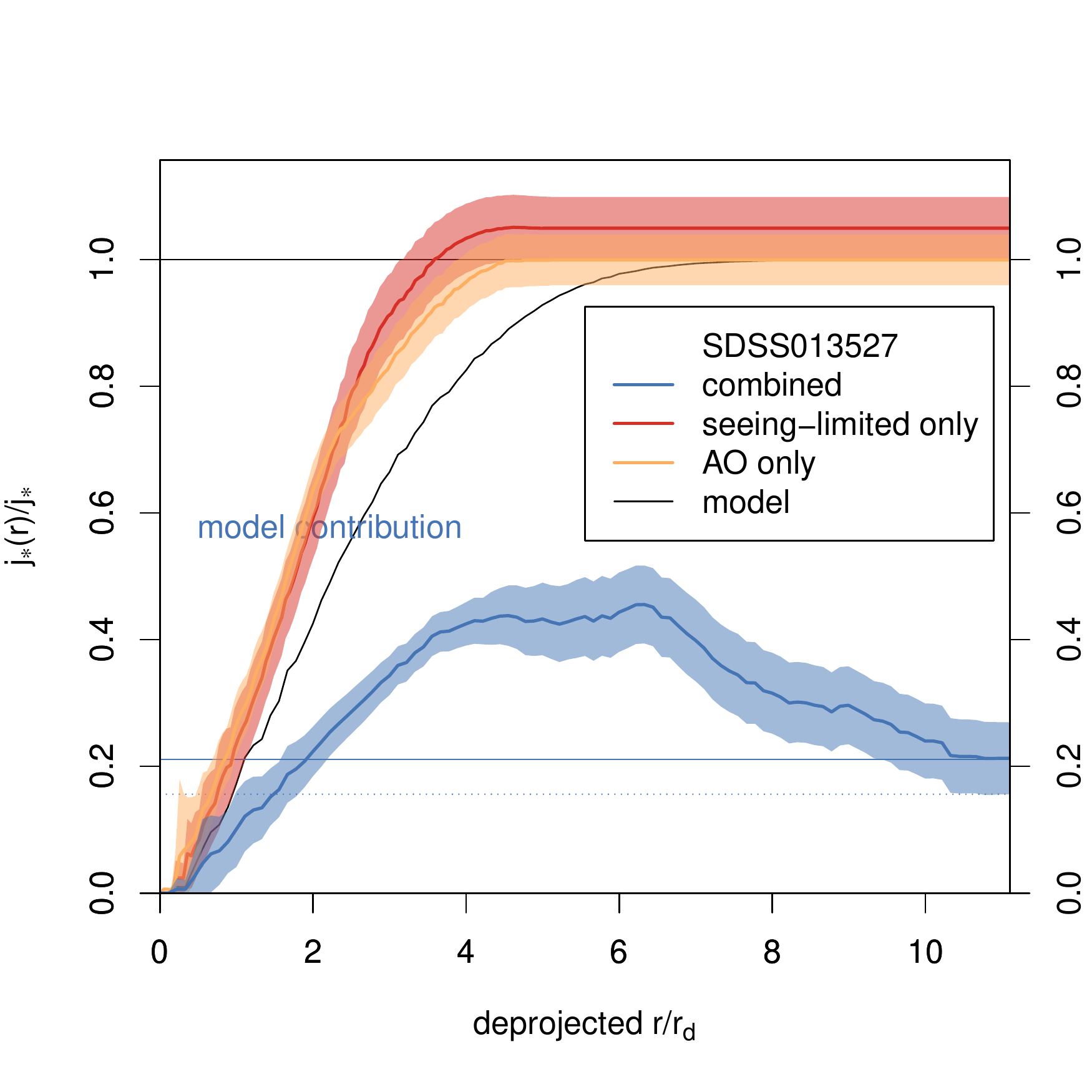}
\includegraphics[width=\columnwidth]{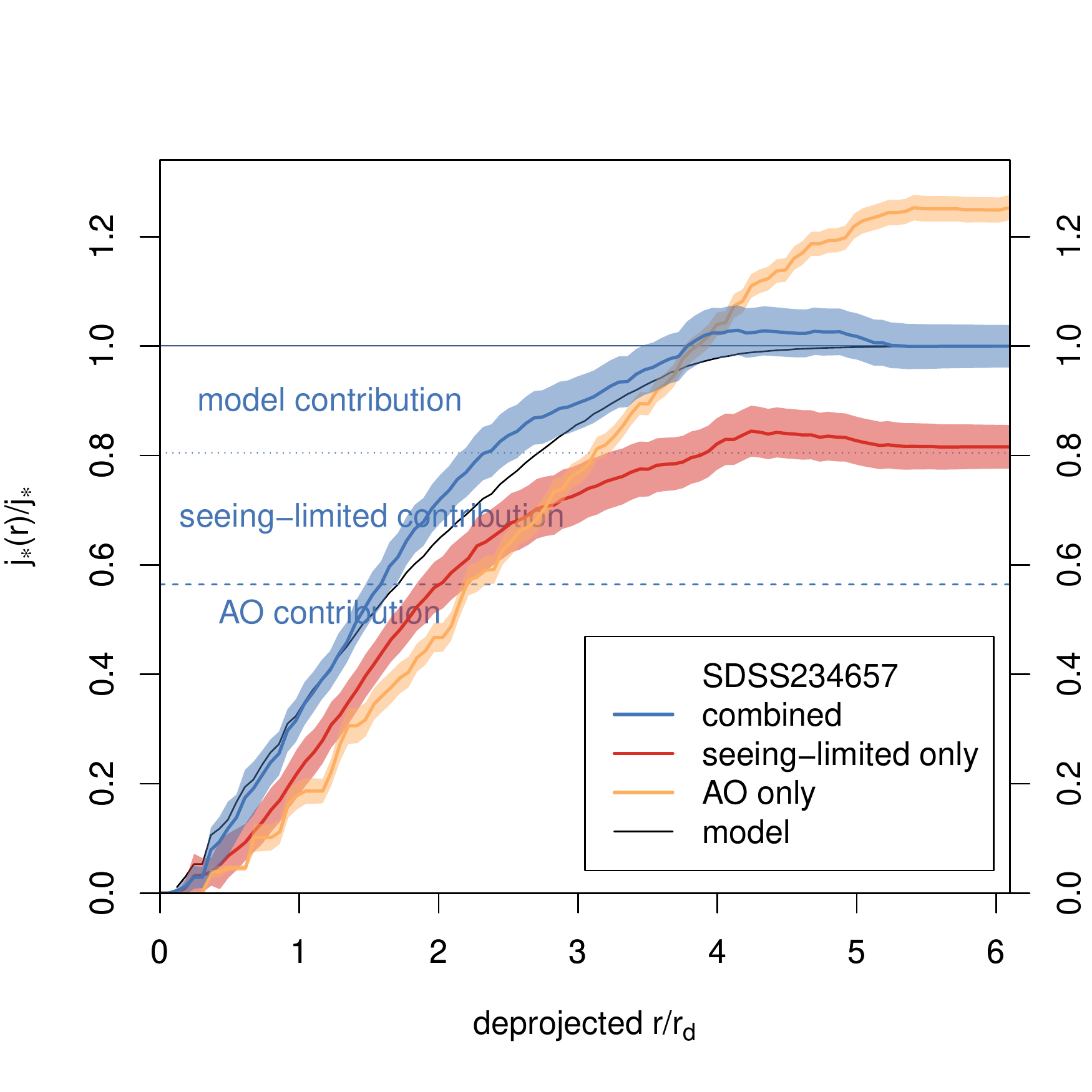}
\includegraphics[width=\columnwidth]{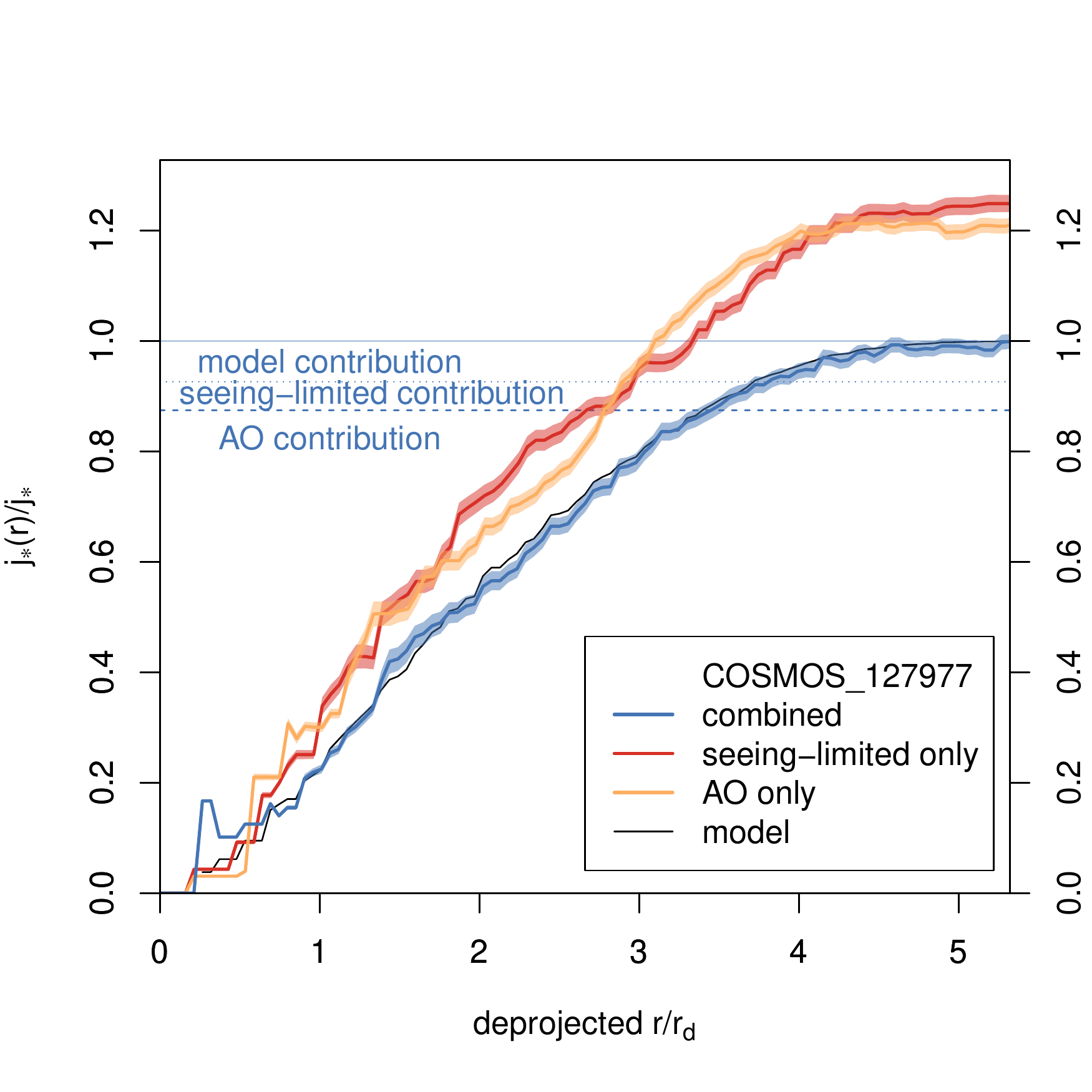}
\includegraphics[width=\columnwidth]{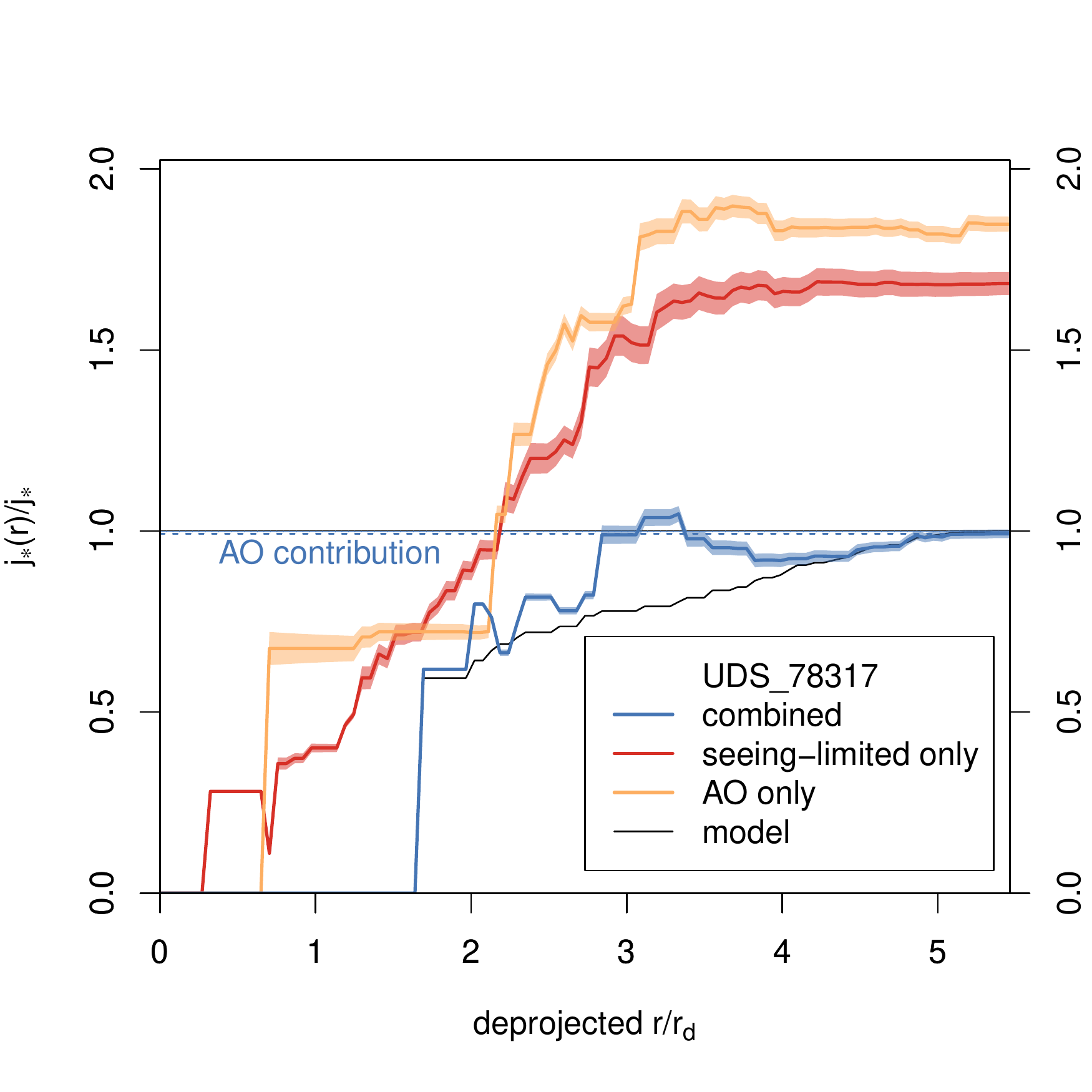}
\caption{(Continued from Figure~\ref{fig:nat_ao}) Cumulative specific AM as a function of radius, showing the contribution from AO data, seeing-limited data and model-informed estimate. Only the galaxies with both natural seeing and AO data are shown in this Figure.}
\label{fig:nat_ao2}
\end{figure*}

This method differs from traditional measurements in the following ways:
\begin{enumerate}
\item Long-slit spectroscopy, such as in RF12, can suffer from misalignment between the kinematic and photometric major axes. The 2D nature of IFS means that the kinematic major axis need not be known {\it a priori} \citep{Sweet+2016}.
\item Most other work, whether long-slit or IFS, utilises the proxy $\tilde{j_*} = krv$, with single fitted characteristic radius $r$ and velocity $v$, and the factor $k = k(n)$ an empirical function of S{\'e}rsic index $n$ in an effort to account for the variation in $j_*$ with morphology. However, the velocity fields of many galaxies may not always be well described by simple 1D rotation curves assumed by this model. OG14 showed for nearby spiral galaxies that spaxel-wise integration of fine spatial resolution IFS data gives an order-of-magnitude improvement in \emph{precision}\footnote{We note that at high redshifts, where physical spatial resolution is coarser, the decrease in uncertainty may not be so dramatic. However, we choose to use this consistent method throughout the paper.}. The corresponding correction 
is described by
\begin{equation}
\left(\frac{j_*}{10^3\ {\rm kpc\ km\ s^{-1}}}\right) \approx 
 1.01\left(\frac{\widetilde{j_*}}{10^3\ {\rm kpc\ km\ s^{-1}}}\right)^{1.3}.
\end{equation}
This is likely to be even more critical for high-$z$ and local clumpy galaxies, such as those presented in this work. In this work we compare the spaxel-wise integration method to the traditional rotation curve model method for local turbulent DYNAMO disk galaxies as well as two $z\sim 1.5$ systems.
\item In the inner regions of the galaxy, where the rotation curve is rapidly changing, beam smearing can cause the velocity field to be underestimated. This is particularly an issue for seeing-limited observations of high-$z$ objects, and can be ameliorated by AO-assisted data of sufficient quality. 
\item Reaching the outskirts of the galaxy is critical to trace the bulk of the AM \citep[e.g. 0.99$j_*$ at 3$r_e$, as in][]{Sweet+2018}. AO-only data often lacks the sensitivity to reach such high multiples of the effective radius, so we mitigate this by including seeing-limited data as well.
\end{enumerate}
We demonstrate (ii), (iii) and (iv) in this paper.\\

The method described above was performed on the two $z\sim 1.5$ galaxies and 20 DYNAMO galaxies described in the previous Section. We discarded eight DYNAMO galaxies where the fit failed due to poor S/N or disturbed kinematics indicative of a merger, in order to obtain a meaningful control sample against which to compare the $z\sim 1.5$ galaxies. Six of the remaining 12 galaxies were observed with both AO and natural seeing, and our measurements for those are presented in Table~\ref{tab:measured}. We take the philosophy that (modulo S/N) more data generally gives a truer result, and adopt the combined seeing-limited and AO measurements in preference over the seeing-limited or AO data alone. {The exceptions are C22-2, where the AO data are too shallow and limited in radial extent to give any improvement over the seeing-limited observations, and SDSS 013527-1039, where the sky is poorly constrained in the natural seeing map.}
The final adopted results for the 12 DYNAMO and two $z\sim 1.5$ systems are presented in Table~\ref{tab:adopted}. Note that four of these DYNAMO galaxies (C22-2, D13-5, G04-1, G20-2) were presented in O15. Since that paper, we have obtained additional observations, and improved our analysis software to better exclude low S/N spaxels and more carefully fix the kinematic centre, inclination and position angle. The $j_*$ values in this work consequently differ from those in O15; they are within errors on average, but individual galaxies differ by between 28 and 69 per cent. Relative uncertainties have decreased from 13.5 per cent to 10 per cent.

\begin{table*}
    \centering
    \caption{Measured properties of DYNAMO and $z=1.5$ galaxies for natural seeing, adaptive optics and combination of natural seeing and adaptive optics.}
    \label{tab:measured}
\begin{tabular}{llrrrrrr}
  \hline
Name & Obs & $z$ & $r_{flat}$ & $v_{flat}$ & $j_*$ & $\Delta j_*$ & $\tilde{j_*}$ \\
& &  & [kpc] & [km s$^{-1}$] & [kpc km s$^{-1}$] & [kpc km s$^{-1}$] & [kpc km s$^{-1}$] \\
(1)  & (2) & (3) & (4) & (5) & (6) & (7) & (8) \\
  \hline
  C22-2 & GMOS & 0.071 & 1.3 & 164 & 449 & 53 & 498 \\
  C22-2 & OSIRIS & 0.071 & 2.5 & 247 & 1101 & 216 & 1281 \\
  C22-2 & GMOS+OSIRIS & 0.071 & 2.5 & 171 & 384 & 39 & 413 \\
  D13-5 & GMOS & 0.075 &  0.5 & 182 & 10343 & 2068 & 10342 \\
  D13-5 & OSIRIS & 0.075 & 0.4 & 174 & 567 & 88 & 565 \\
  D13-5 & GMOS+OSIRIS & 0.075 & 0.4 & 171 & 466 & 51 & 437 \\
  G04-1 & GMOS & 0.130 & 1.5 & 227 & 976 & 124 & 1051 \\
  G04-1 & OSIRIS & 0.130 & 0.9 & 242 & 1065 & 173 & 1119 \\
  G04-1 & GMOS+OSIRIS & 0.130 & 0.7 & 221 & 991 & 118 & 1025 \\
  G20-2 & GMOS & 0.141 & 0.9 & 143 & 295 & 20 & 282 \\
  G20-2 & OSIRIS & 0.141 & 0.6 & 122 & 281 & 29 & 281 \\
  G20-2 & GMOS+OSIRIS & 0.141 & 0.7 & 121 & 309 & 19 & 277 \\
  SDSS 013527$-$1039 & GMOS & 0.127 & 1.3 & 120 & 424 & 55 & 467 \\
  SDSS 013527$-$1039 & OSIRIS & 0.127 & 1.0 & 118 & 404 & 53 & 429 \\
  SDSS 013527$-$1039 & GMOS+OSIRIS & 0.127 & 1.0 & 111 & 85 & 12 & 185 \\
  SDSS 234657+0056 & GMOS & 0.182 & 1.1 & 92 & 330 & 23 & 337 \\
  SDSS 234657+0056 & OSIRIS & 0.182 & 0.9 & 92 & 506 & 84 & 540 \\
  SDSS 234657+0056 & GMOS+OSIRIS & 0.182 & 0.7 & 85 & 404 & 41 & 428 \\
  COSMOS 127977 & KMOS & 1.62 & 3.4 & 215 & 3154 & 434 & 3242 \\
  COSMOS 127977 & OSIRIS & 1.62 & 4.7 & 248 & 3074 & 363 & 3045 \\
  COSMOS 127977 & KMOS+OSIRIS & 1.62 & 3.9 & 236 & 2532 & 227 & 2279 \\
  UDS 78317 & KMOS & 1.47 & 0.4 & 57 & 569 & 76 & 484 \\
  UDS 78317 & OSIRIS & 1.47 & 0.8 & 44 & 628 & 114 & 534 \\
  UDS 78317 & KMOS+OSIRIS & 1.47 & 0.2 & 29 & 338 & 64 & 200 \\
   \hline
\end{tabular}\\
Columns: (1) galaxy identifier;  (2) data source (natural seeing GMOS or KMOS, adaptive optics OSIRIS, or combination of natural seeing and adaptive optics); { (3) redshift;}
 (4) radius at which rotation curve becomes flat; (5) asymptotic velocity; (6) stellar specific AM; (7) measurement uncertainty in $j_*$; (8) stellar specific AM using $\tilde{j_*}=krv$. 
 Note that only the galaxies with both natural and AO are presented in this table.
\end{table*}

\subsection{Bulge-to-total ratios}

We measure bulge-to-total mass ratios $\beta$ for the DYNAMO galaxies using HST imaging where available, and IFS continuum maps otherwise, using the methods described in \citet{Savorgnan+2016}. Briefly, multiple components including bulge, disk, AGN, bar and ring are identified from the images and unsharp masks. Corresponding S{\'e}rsic, exponential, Gaussian, Moffat, Ferrer and symmetric Gaussian ring components are simultaneously fit to the circularized 1D light profile, where the final functional forms are chosen through an iterative process in order to minimise the residuals. 

For the high-$z$ galaxies we use the HST archival imaging to estimate $\beta$ using a similar method to the 2D bulge-disk decompositions described in \citet{fisher2008} and used for the THINGS galaxies that are presented in S18 and included in this work. In both cases the photometry is consistent with a bulgeless galaxy. 

The two methods both include a careful, iterative approach and are not reproducible with automatic routines \citep{fisher2008,Savorgnan+2016}. Despite the differences they have been shown to achieve consistent results for the same galaxies, with the 1D fits undertaken for DYNAMO because of the lower failure rate for decompositions and more instructive isophotal analysis \citep{Savorgnan+2016}. The 2D fits were appropriate for the more distant $z\sim 1.5$ sources with lower physical resolution.

\subsection{Stellar masses}

Our {integrated} stellar mass measurements for the DYNAMO sample are derived from 2MASS $K_s$-band apparent magnitudes (4$R_s$ i.e. 99\% of the light), with Galactic extinction correction according to \citet{Schlafly+2011}. 
We conduct an empirical $k$-correction \citep{Glazebrook+1995} with a universal mass-to-light ratio $M/L_{K_s} = 0.5 M_\odot/L_\odot$ assuming a \citet{Kroupa2001} IMF for consistency with the THINGS, RF12 and CALIFA samples from S18, which we include here as control samples. The typical error introduced by assuming a constant M/L ratio is less than 20 per cent \citep{Bell+2003}. Using stellar masses following the methods in \citet{Kauffmann+2003} instead does not change the results of this paper, but not all of the DYNAMO sample have these available so we opt for the more complete set of measurements.

The $z\sim 1.5$ masses are derived from SED fitting with {\sc hyper-z} \citep{Bolzonella+2000}, using the methods described in \citet{Swinbank+2017}. We adopt measurement uncertainties of a standard 0.2 dex to conservatively account for deviations in results between common SED fitting codes, and possible effects of low photometric signal-to-noise \citep{Mobasher+2015}.

\subsection{Velocity dispersions}
In the $z\sim 1.5$ galaxies the velocity dispersions are calculated as the median of the KMOS observed dispersion map, corrected for instrumental broadening and beam smearing according to the methods of \citet{Johnson+2018}. {Briefly, we create KMOS data cubes for $10{^5}$ model galaxies and convolve these intrinsic data with PSFs typical of the sample to obtain mock observed data. We compare the properties of the intrinsic and mock observed cubes to obtain correction factors that can be applied to the sample.}

For DYNAMO, the velocity dispersions are assumed to be constant across the disk and are derived from fits to the to GMOS observations, as described in \citet{Fisher+2019}.

Velocity dispersions for THINGS are based on CO measurements presented in \citet{Mogotsi+2016}, under the assumption that CO dispersions trace H$\alpha$ dispersions, as discussed in \citet{White+2017} for gas-rich, turbulent DYNAMO galaxies, and demonstrated for one disk galaxy at $z=1.4$ by \citet{Ubler+2018}.

\section{The effect of image quality on measured specific angular momentum}
\label{sec:dataset}

In this Section we discuss the effects of spatial resolution and radial coverage on the determination of angular momentum. We present the 2D maps and 1D rotation curves for the two $z\sim 1.5$ systems, and then give a quantitative comparison of specific AM measured in the three data types (seeing-limited, adaptive optics, combined) as described in Section~\ref{sec:methods}, for the $z\sim 1.5$ COSMOS 129799 galaxy as well as for local turbulent galaxies in the DYNAMO sample. \\

Figures~\ref{fig:cosmos_maps} and~\ref{fig:uds_maps} show resolved maps of H$\alpha$ intensity and velocity for natural seeing KMOS observations and OSIRIS AO-assisted data. For both COSMOS 127977 and UDS 78317, the seeing-limited maps probe to larger radii than the AO maps, while the AO maps display more structure. For COSMOS 127977, the AO H$\alpha$ intensity map confirms the clumpy nature of this galaxy seen in HST imaging. {The difference in substructure between AO and seeing-limited maps is at the scale of the KMOS PSF and can likely be attributed to the difference in PSFs of the two maps (0".1 vs. 0".6). T}he corresponding {AO} velocity map is consistent with the seeing-limited classification of a rotating disk galaxy, but also reveals a kinematic twist along the minor axis that is not seen in natural seeing. In the case of UDS 78313, the finer PSF of the AO data uncovers additional substructure that is not evident in the seeing-limited data. 
\\

The rotation curves presented in Figure~\ref{fig:rot_curves_cos} illustrate the complementarity of both natural seeing observations and AO-assisted data. The natural seeing velocity field of COSMOS 127977 is well fit by the model rotation curve in the top panel of Figure~\ref{fig:rot_curves_cos}, while the middle panel displays the additional structure revealed in the AO maps. Combining the two datasets as described in Section~\ref{sec:methods} leads to a more well-constrained model rotation curve (lower panel), and therefore a more accurate determination of total $j_*$. We reiterate that the model is only used to calculate $j_i$ in the spaxels $i$ where low signal-to-noise prohibits calculation from the data. The 2D spatial structure traced by the integral field data (that is, by the AO-assisted maps in the inner regions of the galaxy and the seeing-limited maps in the outer regions where the AO data has low S/N) is used to calculate $j_i$ elsewhere. The rotation curves for merger system UDS 78317 are in stark contrast to the well-behaved rotation curves of COSMOS 127977. None of the 1D velocity fields are well fit by model rotation curves; for this system there is no clear benefit to using any one of the three datasets in determining $j_*$.\\

Figures~\ref{fig:nat_ao} and~\ref{fig:nat_ao2} show the normalised cumulative stellar specific angular momentum $j_*(\leq r)/j_*$ in order to illustrate the degree of convergence, and the relative contribution to $j_*$ made by the AO-assisted data, the natural seeing data and the model-informed estimate. We include the six DYNAMO and two high-$z$ galaxies for which we have both AO and seeing-limited observations. 
All eight qualitatively show a strong degree of convergence illustrated by the flatness of the cumulative $j_*$ profile, owing to the large multiples of the disk radius $3.5 < r/r_d < 10$ observed. {The horizontal, dashed and dotted lines in these figures correspond to the contribution of the AO, seeing-limited and model spaxels as a fraction of total $j_*$; } on average 69, 18 and 13 per cent respectively. Note that the intersection of the profile with {these lines} does not correspond to a strict radial cut, but can instead be interpreted as an approximate mean boundary to the physical region where the two data types and model \emph{most strongly} contribute.  For example, for C222 the AO-assisted data are most critical for $r/r_d \lesssim 2.75 $, the seeing-limited data between $2.75 \lesssim r/r_d \lesssim 5.5$, and the model-informed estimate beyond $5.5 \lesssim r/r_d$. However, since the length of the $x$-axis $r/r_d$ is set by the limit of the data, the observations contribute at radii as great as $r/r_d \lesssim 7.3 $.\\

\begin{figure}
\includegraphics[width=\columnwidth]{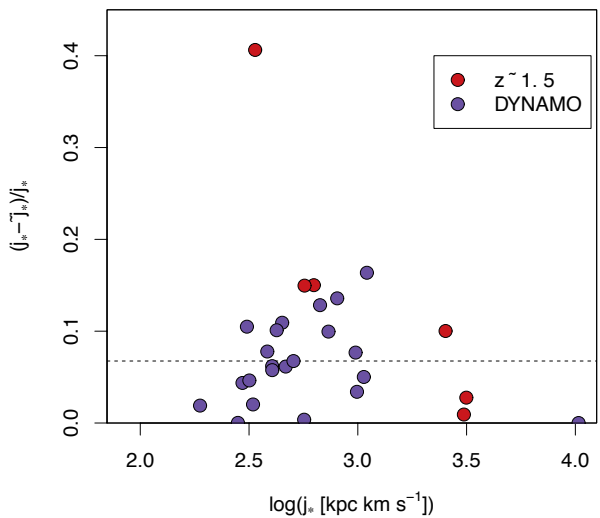}
\caption{Comparison of specific angular momentum measured from spaxel-wise integration ($j_*$) with those measured via the commonly-used rotation curve model method ($\tilde{j_*}$). The { fractional} difference is calculated with $ (j_* - \tilde{j_*})/j_*$. Measurements using all three datasets (AO-assisted, seeing-limited, combination) are shown. The horizontal line represents the mean { fractional} difference.}
\label{fig:j_tot_tilde}
\end{figure}

For comparison with our adopted spaxel-wise integration method we also compute the rotation curve model, adopting $\tilde{j_*} = 2 v_{flat} r_d$, which is commonly used in studies of $j_*$ large samples of galaxies at $z\sim 1.5$ \citep[e.g.][]{Swinbank+2017}. We include these values in Tables~\ref{tab:measured} and~\ref{tab:adopted}. Figure~\ref{fig:j_tot_tilde} illustrates the per cent difference between specific angular momenta measured with the two methods, $\Delta j_* = (j_* - \tilde{j_*})/j_*$. Across all datasets, $\tilde{j_*}$ differs from our two-dimensional integrated $j_*$ by 6.75 per cent, which is less than the level of measurement uncertainty. The cases where $\tilde{j_*}$ differs from our $j_*$ by more than one standard deviation are the targets that have particularly clumpy morphology (e.g. G08-5, G20-2, SDSS 033244+0056) or where the kinematic map shows substructure (e.g. UDS 78317, which has the highest $\Delta j_* = 0.406$ for the combined AO + seeing-limited dataset). Our primary motivation for adopting the spaxel-wise integration method described above is to account for the diversity of galaxies such as these. The result can be interpreted to mean that $j_*$ measured from spaxel-wise integration and $\tilde{j_*}$ measured from the rotation curve model method in general give consistent results.\\

We now consider the systematic effects of the three different data sets (GMOS, OSIRIS, KMOS as well as DYNAMO versus $z\sim 1.5$ galaxies). As discussed above these different sets vary in both radial coverage of the velocity profile and spatial resolution, which can be important for tracing the bulk of AM. We will consider both the effect on measurement uncertainties and the systematic effects on the value of total $j_*$ across data type. In Table~\ref{tab:nat_ao} we compare the three datasets for COSMOS~127977,~UDS 78317 and the six DYNAMO galaxies for which we have both natural seeing and AO data.

Overall there is a general decrease in median measurement uncertainty $\Delta j_*$ as more information is included in the fit, as one would expect for random noise. For COSMOS 127977 the measurement uncertainty decreases from 14\% with KMOS-only to 9\% when the OSIRIS and KMOS data is combined. However, in UDS~78317 the measurement uncertainty increases from 13\% for KMOS-only to 18\% for the combined measurement. This is due to additional substructure being detected with OSIRIS, as illustrated in Figures~\ref{fig:cosmos_maps} and~\ref{fig:uds_maps}. 

The DYNAMO sample median uncertainty on $j_*$ decreases from 13\% (GMOS-only) or 16\% (OSIRIS) to  10\% for the combined data set. The higher median uncertainty for AO than natural seeing in DYNAMO could be an artefact of finer resolution of clumps\footnote{This is assuming that the clumps have their own velocity field which disturbs the velocity field of the galaxy \citep[as seen in simulations e.g.][but not yet seen in observations]{Ceverino+2012}.}, and lower signal-to-noise at high multiples of $r_e$, both of which mean the model fit is less certain, which in turn contributes to uncertainty in $j_*$. 

Our philosophy to adopt {in general} the dataset with the most data results in selecting the dataset that has the lowest relative measurement uncertainty on $j_*$. {For C22-2 and SDSS 013527-1039, where we respectively adopt seeing-limited and AO-only data, the relative measurement on $j_*$ is 1.7 and 0.2 per cent larger respectively.}

Using the OSIRIS data on its own can sometimes give dramatically different values of $j_*$, due to the more restricted coverage of the velocity field compared with GMOS or KMOS. For example, UDS 78317 $j_*$ measured with OSIRIS is a factor of two higher than the combined KMOS+OSIRIS dataset. 

The KMOS-only measurements at $z\sim 1.5$ in general disagree with the combined KMOS+OSIRIS by at least 20$\%$. 
For UDS 78317 the disagreement in $j_*$ is at the 60 per cent level. We note that the seeing-limited data are consistent with a rotating disk, but the H$\alpha$ intensity maps and `multi-polar' velocity maps together indicate that this may be two rotating galaxies undergoing a merger. It is possible that using kinemetry \citep{Krajnovic+2006} or modelling \citep{Rodrigues+2017} may be able to identify this system as a merger in the KMOS data, however this is beyond the scope of this paper.  The implication from our detailed observations of these two $z \sim 1.5$ galaxies is that the systematic uncertainty on $j_*$ is at best 20 percent, and in some cases may be significantly higher. While we acknowledge that our sample is small, nonetheless we recommend that caution should be taken when interpreting $j_*$ that is measured with only natural seeing at $z>1$. 

The GMOS-only measurements can be thought of as analogous to the KMOS+OSIRIS dataset at $z\sim 1.5$. Both AO enabled observation with OSIRIS at $z\sim 1.5$ and the seeing-limited ($\sim$0.5-0.7 arcsec) observations of DYNAMO galaxies with GMOS offer a resolution in the central part of the rotation curve of order $\sim$1~kpc. Moreover, both data sets reach sufficiently far in radius to adequately constrain the flat rotation curve. Observations of DYNAMO galaxies that include OSIRIS offer finer spatial resolution that what is available on unlensed galaxies at $z\gtrsim 1$. We can therefore use this comparison to understand what information is lost on measurements of $j_*$ on $z\gtrsim 1$ galaxies. 

With the exception of one target (D13-5), the $j_*$ values of GMOS-only observations of DYNAMO galaxies agree with the corresponding GMOS+OSIRIS values to the 11 per cent level. The values of $j_*$ for D13-5 are a factor of two higher in the GMOS-only data than in the combined GMOS+OSIRIS data. This discrepancy arises because the seeing-limited data are not well fit by the exponential disk model used for extrapolation. 
We note that this measurement has the highest measurement uncertainty ($20\%$) in our sample. When combining the data, the low S/N GMOS spaxels are replaced by OSIRIS, so the combination is well fit. We interpret this to indicate that the general correspondence between GMOS-only and GMOS+OSIRIS in DYNAMO galaxies indicates that using KMOS+OSIRIS at $z\sim 1.5$ is sufficient to achieve a robust result in that using finer spatial resolution would not appreciably alter the measurement of $j_*$.
\\

\begin{table}
    \centering
    \caption{Median $j_*$ and uncertainty in $j_*$ for natural, adaptive optics, and natural + adaptive optics data.}
    \label{tab:comp}
\begin{tabular}{llll}
  \hline
& Natural seeing & adaptive optics & natural + AO \\ 
  \hline
  \hline
$z\sim 1.5$\\
\hline
log($j_*$) & 3.50 $\pm$ 0.06 & 3.49 $\pm$ 0.05 & 3.4 $\pm$ 0.04 \\
$\Delta j_* / j_*$ & 0.14  & 0.12 & 0.09 \\
\hline
\hline
  $z\sim 0$\\
  \hline
log($j_*$) & 2.64 $\pm$ 0.24 & 2.73 $\pm$ 0.10 & 2.60 $\pm$ 0.14 \\
$\Delta j_* / j_*$ & 0.13 $\pm$ 0.02 & 0.16 $\pm$ 0.01 & 0.10 $\pm$ 0.01\\
   \hline
\end{tabular}\\
\label{tab:nat_ao}
\end{table}

The consequence is that high-redshift studies that utilise only natural seeing observations, or only adaptive optics-assisted data, { in general} are likely to measure less well-constrained $j_*$ than if using a combination of seeing-limited and AO-assisted data, and may also marginally overestimate $j_*$. { One might expect that effect of using only natural seeing data would be to \emph{underestimate} $j_*$ due to beam smearing, but since our natural seeing data are scaled for beam smearing we do not see that effect here. }

\section{The relation between stellar mass, specific angular momentum and morphology} \label{sec:bt_jm}

\begin{figure}
    \includegraphics[width=\linewidth]{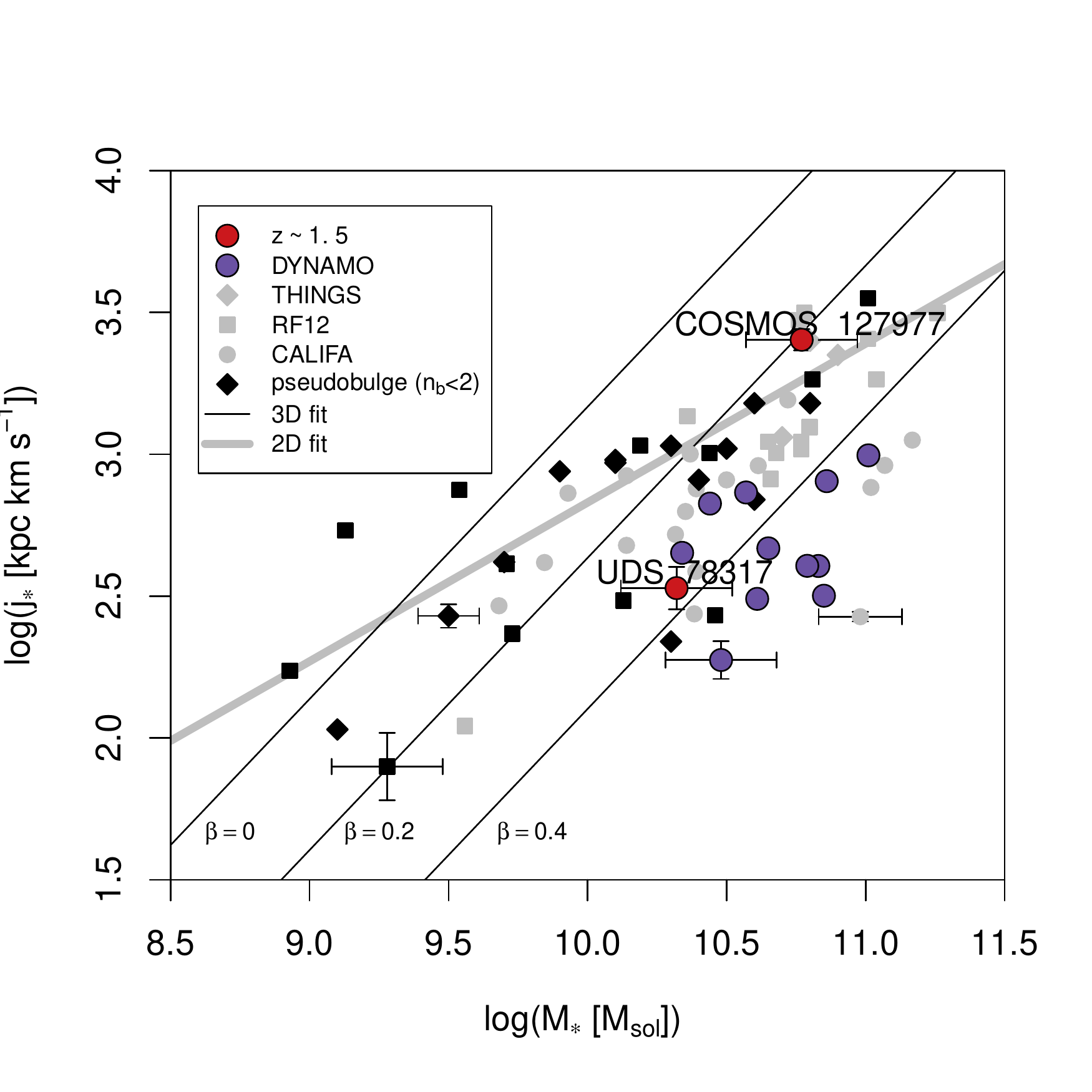}
    \caption{The relation between stellar mass $M_*$ and stellar specific angular momentum $j_*$ for the two $z\sim 1.5$ galaxies (red filled circles) and the local analogue DYNAMO galaxies (purple filled circles), overlaid on the results for normal local disk galaxies (grey and black diamonds, squares and small circles) presented in \citet{Sweet+2018}. The DYNAMO galaxies occupy the low-$j_*$ region occupied by moderately bulge-dominated local galaxies, but themselves have low bulge fraction $\beta$. COSMOS 127977 has a high $j_*$ for its stellar mass and is consistent with local spirals. UDS 78317 lies within the DYNAMO scatter, but the OSIRIS observations reveal that this system is a merger.}
    \label{fig:m_j}
\end{figure}

\begin{figure}
    \includegraphics[width=\linewidth]{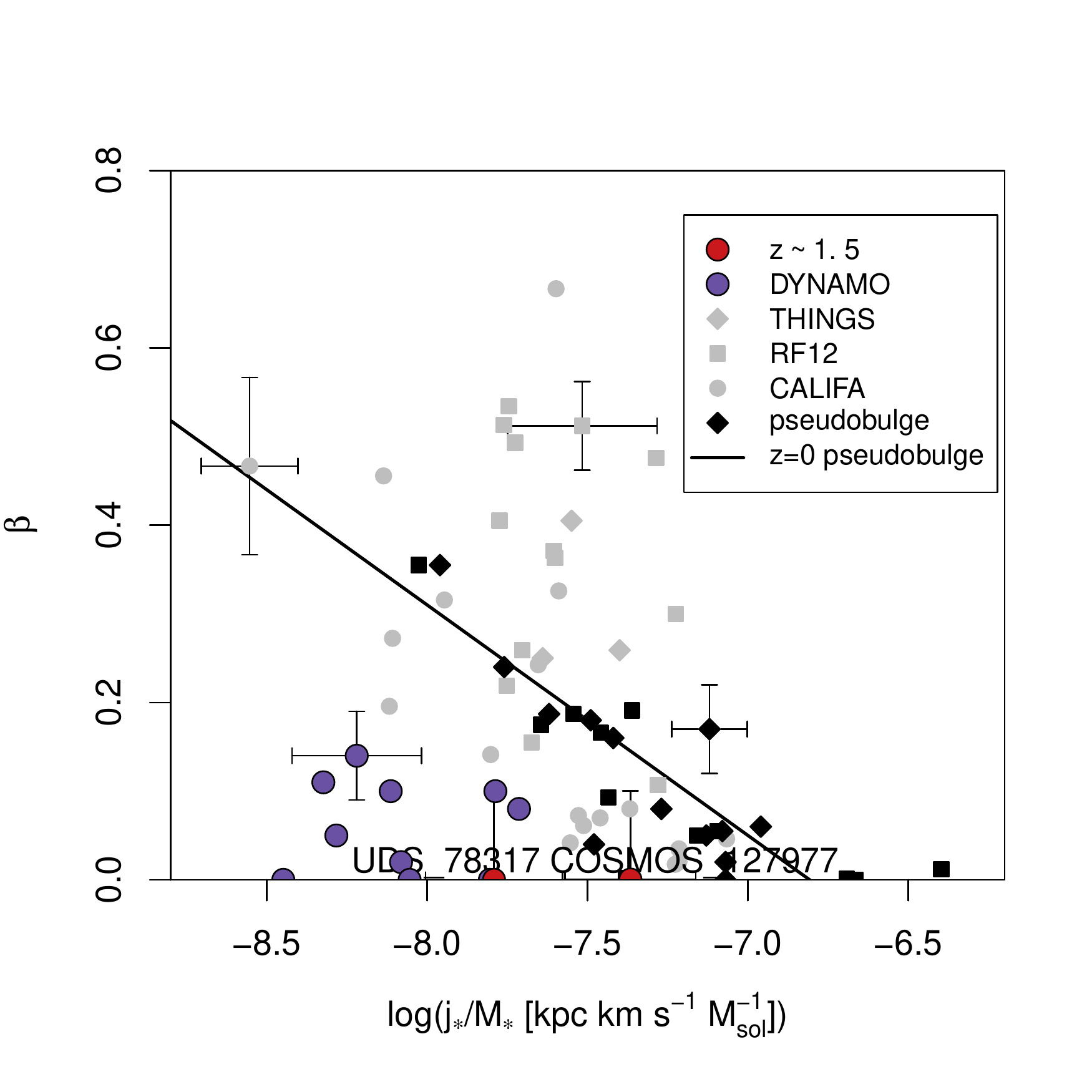}
    \caption{The relation between bulge fraction $\beta$ and stellar specific angular momentum per unit stellar mass $j_*/M_*$ for the two $z \sim 1.5$ galaxies and the local analogue DYNAMO galaxies, overlaid on the results for normal local disk galaxies presented in \citet{Sweet+2018}. The DYNAMO galaxies are offset from the relation defined by local galaxies that host pseudobulges, with relatively small bulges for their $j_*/M_*$ ratios. COSMOS 127977 is also below the local relation, but within the dispersion of the local sample. UDS 78317 is consistent with the highest $j_*/M_*$ DYNAMO galaxies.}
    \label{fig:bt_jm}
\end{figure}

\begin{figure}
    \includegraphics[width=\linewidth]{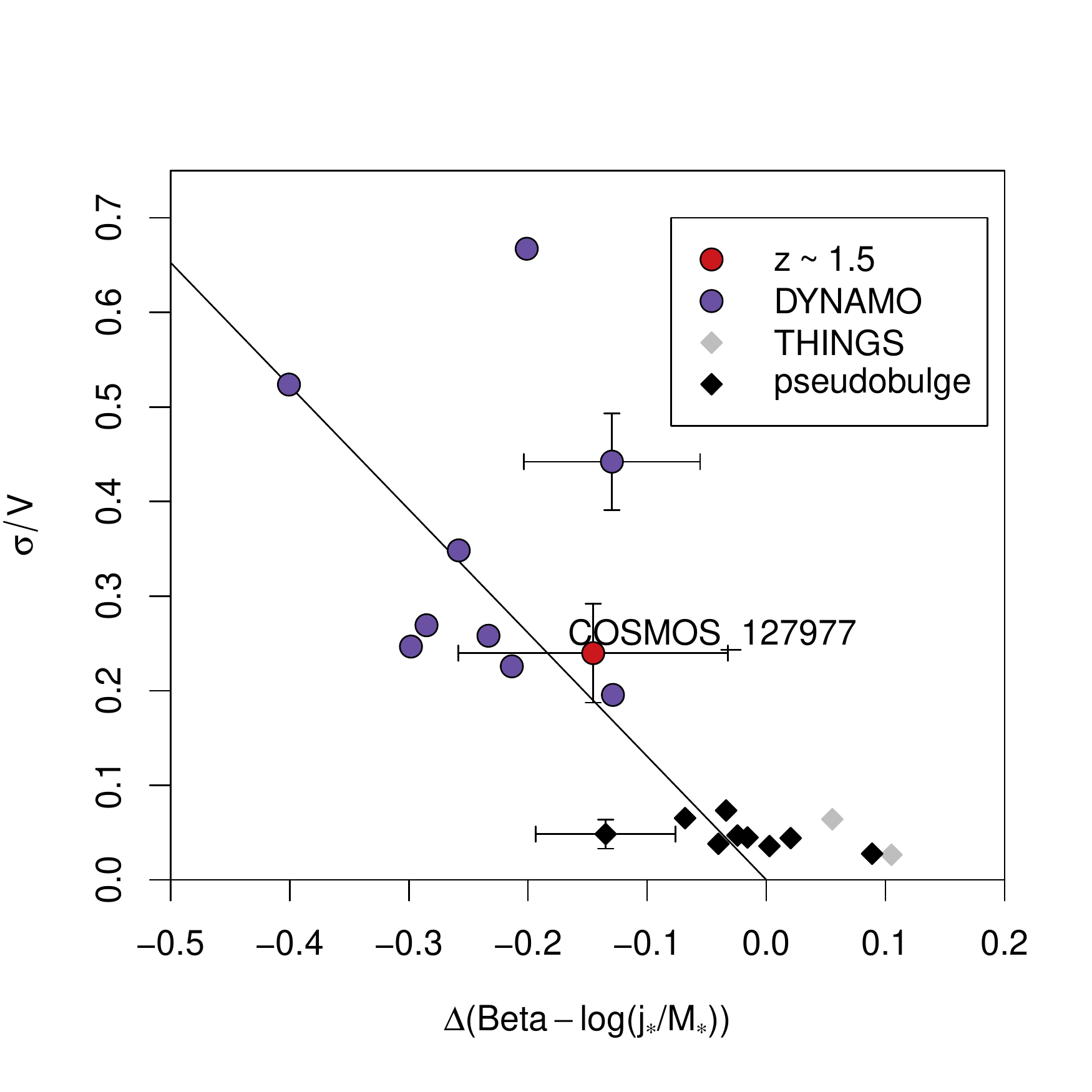}
    \caption{The relation between the offset to $\beta-j_*/M_*$ relation and ratio of velocity dispersion { to rotational velocity} $\sigma/v_{flat}$ for the $z\sim 1.5$ galaxy COSMOS 127977 and the local analogue DYNAMO galaxies, overlaid on the results for normal local disk galaxies THINGS presented in \citet{Sweet+2018}. 
    DYNAMO galaxies and COSMOS 127977 are offset from THINGS galaxies. 
    }
    \label{fig:dbtjm_vsigma}
\end{figure}

In this Section we analyse the relationship between stellar mass, specific angular momentum (AM) and morphology of $z\sim 1.5$ galaxies and their local analogues in the DYNAMO sample, compared with normal local galaxies from THINGS, CALIFA and RF12 as presented in \citet{Sweet+2018}, in order to quantify how bulge growth and disk stability are correlated with the build-up of AM and stellar mass $M_*$. We use our full DYNAMO sample (that is, we include the galaxies for which we have only seeing-limited data, those for which we only have AO-assisted data, and those for which we have both) since there is likely to be no significant systematic effect on stellar specific AM $j_*$ as demonstrated in Section~\ref{sec:dataset}.\\

The $M_*-j_*$ plane is shown in Figure~\ref{fig:m_j}. DYNAMO galaxies all lie below the 2D relation for normal local disk galaxies, in the region occupied by normal galaxies with moderate ($\beta \sim 0.4$) bulge fraction. There is a large amount of scatter, likely due to the range of specific star formation rates and consequent range of star formation-induced turbulence, which is connected via disk stability to a corresponding range in $j_*$. The $z\sim 1.5$ disk galaxy COSMOS 127977 is consistent with local $L_*$ galaxies, with a large $j_* = 2500 \pm 200$ kpc km s$^{-1}$ and log$(M_*/M_\odot) = 10.77$. The merging system at $z\sim 1.5$, UDS 78317, lies below the $M_*-j_*$ relation amongst the DYNAMO sample. We point out that, since the system is not in dynamical equilibrium, the assumptions made in determining $j_*$ in this manner are not valid, so this measurement may not be meaningful.\\

Motivated by the empirical finding by OG14 and S18 that the 3D fit \begin{equation}
\label{eqn:OG14-10}
\frac{j_*}{10^3\ {\rm kpc\ km\ s^{-1}}} = 
k e^{(-g\beta)} \left(\frac{M_*}{10^{10}M_\odot}\right)^\alpha,
\end{equation} yields $\alpha \sim 1$, and the physical interpretation that $j_*/M_* = J_*/M_*^2$ is connected with disk stability \citep{OG14}, we show the $\beta-j_*/M_*$ plane in Figure~\ref{fig:bt_jm}. This Figure illustrates that the location of UDS 78317 and the DYNAMO galaxies on Figure~\ref{fig:m_j} cannot simply be explained by high $\beta$, since these systems do not have large bulges. They all have low $\beta$ for their ratio of $j_*$ to $M_*$ compared with the relation for normal local disk galaxies that host pseudobulges, presented in \citet{Sweet+2018}. COSMOS 127977 is within the scatter of the local control samples owing to its high $j_*$. {If our assumption that the gas and stars corotate is incorrect for this $z\sim 1.5$ galaxy as discussed in Footnote~\ref{foot:corotation}, then its log($j_*/M_*$) would be overestimated by 0.1 dex, moving it marginally below the scatter of the local control samples but still above the turbulent DYNAMO systems.}\\

We investigate the relation between degree of turbulence and distance from the $\beta-j_*/M_*$ relation along the $\beta$ axis in Figure~\ref{fig:dbtjm_vsigma}. The degree of turbulence is quantified by the ratio of velocity dispersion $\sigma$ to rotational velocity $v_{flat}$, which is shown to correlate with clump size in \citet{Fisher+2017}. It stands to reason that clump size correlates with mass as clumps have relatively constant surface brightness both in H$\alpha$ emission \citep{Fisher+2017b} and stellar mass \citep{Cava+2018}. In this sense $\sigma/v_{flat}$ is an accessible proxy for clumpiness. 
There are two separate groups on this plot, where turbulent DYNAMO and $z\sim 1.5$ galaxy COSMOS 127977 have a larger offset from the $\beta-j_*/M_*$ relation than regular local THINGS galaxies, which have low dispersion. $\sigma/v_{flat}$ for UDS 78317 is too high for that galaxy to appear on this figure, but this is not unexpected, since as a merging system its $v_{flat}$ does not correspond to that of a rotating disk.
The implied result of the broad trend between clumpiness and offset from $\beta-j_*/M_*$ is that galaxies that deviate further from the pseudobulge relation are still in the process of building their bulges. 
Interestingly, high-$z$ galaxy COSMOS 127977 is consistent with local analogue DYNAMO galaxies only once its turbulence is accounted for, suggesting that ignoring this parameter can yield an incomplete picture of galaxy evolution.

\section{Discussion}
\label{sec:discussion}

In this Section we discuss the potential evolution of DYNAMO galaxies in $M_*-j_*-\beta$ space, and compare the resulting implications for high-$z$ galaxies with our findings at $z\sim 1.5$.

We have seen in the previous Section that DYNAMO galaxies have low $j_*$ for a given stellar mass, occupying the space generally populated by early-type galaxies. 
However, unlike early types, DYNAMO galaxies do not have a large central bulge and are clumpy and turbulent, representing an earlier stage of evolution than typical local disk galaxies. 
The DYNAMO galaxies exhibit a large scatter, but this is not due to the inclusion of { present-day} mergers, since this sample excludes systems that do not appear to be disks (i.e., those with disturbed velocity maps and/or non-exponential surface brightness profiles)\footnote{
There may still be remnants of past mergers in the sample, since past gas-rich mergers would leave disky, bulge-less systems \citep{Hopkins+2009}.}.
The scatter can be at least partially attributed to the range in star-formation properties, noting that they have specific star formation rates consistent with galaxies between $0\lesssim z\lesssim 2$ (e.g. C22-2 is consistent with other $z = 0$ galaxies).
If the clumps in these galaxies remain bound and migrate to the centre, as proposed in \citet{Ceverino+2012}, then perhaps the mass of the clumps could build the bulge mass, and these high-$z$ analogues may evolve to reach the present day relation traced by pseudobulges in Figure~\ref{fig:bt_jm}. 
For three of these targets (D13-5, G04-1 and G20-2), data presented in \citet{Fisher+2017} can be used to make a back-of-envelope calculation of future bulge fraction $\beta_{future} = (M_{clump} + \beta M_*)/(M_* + M_{clump})$, where $M_{clump} = SFR_{clump} t_{dep}$ is the clump mass available to build the bulge, $SFR_{clump}$ is the total star formation rate in the clumps, $t_{dep}$ is the depletion time for the galaxy, $\beta$ is the current bulge fraction and $M_*$ is the current stellar mass of the galaxy. The future bulge fractions projected in this way for D13-5, G04-1 and G20-2 are $\beta_{future}$ = 0.19, 0.24 and 0.25 respectively. These estimates move these three galaxies into the range of the pseudobulge relation in Figure~\ref{fig:bt_jm}, indicating that the clumps could contribute to building up the bulges, though perhaps not on their own. Additionally, if there was future gas accretion there would need to be subsequent secular evolution to keep these galaxies on the pseudobulge relation. We have a current (Cycle 25) HST program (PI: Fisher) to measure clump stellar masses, which will improve these estimates. In a future paper we will investigate whether or not summing the mass of the clumps with the bulge is sufficient to relocate DYNAMO galaxies to the pseudobulge relation, or whether there is some additional mechanism required.

Now, the distance to the pseudobulge relation is generally correlated with degree of turbulence, quantified as $\sigma/v_{flat}$, where $\sigma$ is not from thermal pressure alone but also from turbulence pressure due to star formation. In the case of DYNAMO galaxies, these gas-rich, turbulent disks represent an earlier stage in evolution and have not yet built up their bulges. The contribution from star formation-induced turbulence is high owing to the large clumps and high star formation rates, so $\sigma/v_{flat}$ is a proxy for clumpiness. As the DYNAMO galaxies evolve towards the pseudobulge track, they move to the right of Figure~\ref{fig:dbtjm_vsigma}, so they must also decrease their $\sigma/v_{flat}$ in order to remain consistent with the trend shared with normal local galaxies. This would require star formation to decrease and the disk to settle as they move closer to the pseudobulge track and become more like typical present-day disk galaxies. The corollary is that the pseudobulge relation breaks down for clumpy galaxies, and should not be used at $z \gtrsim 1$.


There is still the outstanding question as to why high-$z$-like disks exist today, since their clumpy nature and low AM do not necessarily follow from their high star formation rates by which they were selected. They may have formed in the high-redshift universe and have somehow survived in their clumpy, bulgeless form to the present day, evolving less than their counterparts owing to a relatively under-dense environment. Alternatively they may have formed more recently, but in unique environments where the conditions resemble those in the denser, earlier universe. Analysis of their environments and stellar ages may help to distinguish between these scenarios.\\



If we assume that DYNAMO galaxies are in fact analogues of galaxies at $0\lesssim z \lesssim 2$, then comparing to typical local galaxies in THINGS can be informative about the redshift evolution of $j_*$. For DYNAMO in this work and O15 we see low $j_*/M_*$ \footnote{We remind the reader that analysing this ratio (rather than $j_*$, or $j_*/M_*^{2/3}$) is motivated by OG14 and S18, who find that the slope of the $M_*-j_*$ relation $\alpha \sim 1$ for fixed $\beta$; OG14 make the physical interpretation that $j_*/M_* = J_*/M_*^2$ is connected with disk stability.} and low $\beta$, suggesting that disks at $z\sim 1.5$ are likewise in general less stable than at $z\sim 0$. The enhanced star formation-induced turbulence in DYNAMO -- which (together with gas fraction, high dispersion, disky nature, compactness and clumpiness) earns that sample the `high-$z$ analogue' label, since it is also the case in high-$z$ disks -- is consistent with this picture.

It is instructive to confirm the above discussion, where we treat DYNAMO galaxies as local analogues of clumpy high-$z$ disks, with high-quality observations of systems at $z\sim 1.5$. 
Interestingly, neither of the two $z\sim 1.5$ galaxies presented here matches this description. COSMOS 127977 is consistent with typical $z\sim 0$ disks in $M_*-j_*$ and $\beta-j_*/M_*$ space, even though it has an enhanced $\sigma/v_{flat}$ compared with local THINGS galaxies by virtue of its bright, star-forming clumps. Inclusion of the $\sigma/v_{flat}$ turbulence parameter is necessary to see its expected correspondence with local analogues, suggesting that this is an important parameter in understanding galaxy evolution. We note that COSMOS 127977 was selected for observation based on the seeing-limited KGES data. The KGES $z\sim 1.5$ sample exhibits a wide range of rotational velocity and angular size. For this pilot work we pre-selected galaxies 1) that show evidence of rotation in KGES, to ensure a high-quality $j_*$ measurement despite the challenges of observing at such high redshift; and 2) that maximally occupy half of the OSIRIS field of view, to facilitate on-detector beam-switching sky subtraction and sampling with as many resolution elements as possible. This selection is consequently biased towards higher $j_*$ and larger effective radius, thus also to higher $M_*$. COSMOS 127977 may be a more evolved system than most $z\sim 1.5$ galaxies, in the sense that it has experienced the right conditions for its disk to settle and $j_*$ to build up, bringing it nearer the pseudobulge relation.

The other $z\sim 1.5$ system in our sample, UDS 78317, appears to be a merging system, so is not a normal $z\sim 1.5$ disk either, even though it is consistent with local analogues of high-$z$ galaxies in terms of $j_*$, $M_*$, $\beta$ and $\sigma/v_{flat}$, and would seem to confirm the above evolutionary discussion. This raises important points about the effect of image quality on merger / disk interpretation and subsequent ill-advised inclusion of mergers as though they were rotating disks. UDS 78317 appears to be a rotating disk in the seeing-limited data, with the intensity and velocity maps showing no obvious sign that this system may be a merger (excluding more detailed analysis such as kinemetry). 
It is only with the enhanced PSF of the AO maps, which reveal that the system is clumpy and disturbed, that one realises that UDS 78317 may in fact be a merging system and cannot be treated as a rotating disk galaxy. 
This is supported by the SINS/zC-SINF AO survey \citep{ForsterSchreiber+2018}, who made a comparison between deep AO and non-AO data for 34 galaxies at $z\sim 2$. They found that the larger (angular size) sources were broadly consistent between the two datasets, but that for 14 of the 17 smaller, less resolved sources, the AO maps tended to either further resolve clumps (in six objects) or resolve new structure (in eight objects), including minor mergers in three cases. \citet{Rodrigues+2017} found this to be more serious for KMOS3D at $z\sim 1$, with mergers being misclassified as rotating disk galaxies in 50 per cent of cases. 
We note that gas-rich mergers at high redshift have been shown to evolve to resemble disk-like systems at late times, with a range of final AM depending on the AM vectors of the merging components \citep{Robertson+2006}. We suggest that some fraction of the scatter in $M_*-j_*$ presented by other studies at high redshift 
may thus be driven by including { a greater number of} mergers as if they were rotating disks; this could be checked by measuring $j_*$ in the manner described in the current paper for the SINS/zC-SINF AO and non-AO datasets.
Most other high-$z$ studies \citep{ForsterSchreiber+2006,Burkert+2016,Contini+2016,Swinbank+2017,Harrison+2017} find that $j_* \propto M^{2/3}$ for the 2D relation (with the exception of \citet{Alcorn+2018} who found a shallower slope, but note that they compute $j_*$ from integrated spectra). However, the normalisation of this relation leads to a wide range of conclusions about the redshift evolution $(1+z)^n$, ranging from $n=0$ \citep{Burkert+2016,Alcorn+2018} to $n=-1.5$ \citep{ForsterSchreiber+2006}. Some of this variation may be due to the inclusion of { current} mergers, and other sample selection differences. We also note that these works assume a smooth disk with simple model $j = krv$ instead of utilising the spatially-resolved, 2D maps as we do in this paper. If we perform a similar calculation adopting $\tilde{j_*} = 2 v_{flat} r_d$ assuming pure disks, then $\tilde{j_*}$ is lower than our integrated $j_*$ by 11 and 40 per cent for COSMOS 127977 and UDS 78317 respectively.
This is likely to further increase the scatter in $j_*/M_*$ in those samples \citep[also see][]{OG14}, and may also affect the normalisation.

We are gathering a larger, more representative sample of natural+AO observations of $z\sim 1.5$ disks to quantify the location of high-$z$ galaxies in $M_*-j_*$ and $\beta-j_*/M_*$ space in an accurate, self-consistent manner. \\


\section{Conclusions}
\label{sec:conclusion}

In this paper we have presented high-quality specific AM measurements for local turbulent galaxies in DYNAMO and two $z\sim 1.5$ systems, using a novel combination of AO-assisted and seeing-limited data. We make the following points.

\begin{enumerate}
\item Image quality affects specific AM, in that combining fine PSF of AO-assisted observations in the central regions with high signal-to-noise of seeing-limited data in the outskirts leads to a more well-constrained $j_*$. The mean measurement uncertainty $\Delta j_*/j_*$ is reduced from 13 per cent with seeing-limited data or 16 per cent with AO alone to 10 per cent in the combination of the two data types. The high-quality $j_*$ measured in this manner may be marginally lower than $j_*$ measured with just one type of data or the other.

\item In particular, high-$z$ galaxies observed only in natural seeing may be misclassified as disk galaxies when they are in fact merging systems. Such systems may appear to be consistent with local analogues of high-$z$ galaxies and with theoretical expectations, but cannot be sensibly compared with them. Some of the scatter in $M_*-j_*$ and $\beta-j_*/M_*$ space may be driven by inclusion of merging systems as though they were rotating disk galaxies.

\item Local analogues of high-$z$ galaxies have low $j_*$ for their $M_*$, but also lie below the $\beta-j_*/M_*$ relation for normal local galaxies that host pseudobulges of \citet{Sweet+2018}. Their offset from that relation is broadly correlated with and possibly explained by a physical model whereby enhanced $\sigma/v_{flat}$ is contributed by star formation-induced turbulence.

\item COSMOS 127977, a disk galaxy at $z\sim 1.5$, is consistent with normal local disk galaxies in terms of $j_*$, $M_*$ and $\beta$, albeit with enhanced $\sigma/v_{flat}$. It may be a more evolved system than typical disks at $z\sim 1.5$ and represent an intermediate phase between low-$j_*$ turbulent disks and today's high-$j_*$, smooth galaxies.
\end{enumerate}

In future papers we will extend this work to a larger sample of high-$z$ galaxies and present detailed analyses of spatially-resolved PDF($j_*$) (Gillman et al., in prep., Sweet et al., in prep.).

\section*{Acknowledgements}
{We thank the anonymous reviewer for comments which helped to improve the manuscript.}
DBF, KG acknowledge support from Australian Research
Council (ARC) Discovery Program (DP) grant DP130101460. 
DBF, KG, DO, LW and SMS acknowledge support from ARC DP grant DP160102235.
DBF acknowledges support from ARC  Future  Fellowship FT170100376. 
SG acknowledges the support of the Science and Technology
Facilities Council through grant ST/N50404X/1.
ALT acknowledges support from STFC grant ST/L00075X/1. 
ALT, AMS, RB \& RMS acknowledge support from STFC grant ST/P000541/1. CL has received funding from a Discovery Early Career Researcher Award (DE150100618). KG, CL and SMS have received funding from the ARC Centre of
Excellence for All Sky Astrophysics in 3 Dimensions (ASTRO 3D), through project number CE170100013.  Some of the data presented herein were obtained at the W. M. Keck Observatory, which is operated as a scientific partnership among the California Institute of Technology, the University of California and the National Aeronautics and Space Administration. The Observatory was made possible by the generous financial support of the W. M. Keck Foundation.

\begin{table*}
\centering
\caption{Adopted properties of DYNAMO and $z\sim1.5$ galaxies.} 
\label{tab:adopted}
\begin{tabular}{lllllrrrr}
  \hline
Name & RA & Dec & $z$ & Obs & $M_*$ & $\Delta M_*/M_*$ &  $\beta$ & $\Delta \beta$  \\ 
     & [hms] & [dms] & & & [log($M_\odot$)] & [dex]&   &    \\
(1)  & (2) & (3) & (4) & (5) & (6) & (7) & (8) & (9) \\
  \hline
  C22-2 & 22:39:49.34 & $-$08:04:18.0 & 0.071 & GMOS & 10.34 & 0.20 & 0.10 & 0.05 \\ 
  D13-5 & 13:30:07.01 & +00:31:53.2 & 0.075 & GMOS+OSIRIS & 10.65 & 0.20 & 0.02 & 0.05  \\ 
  G04-1 & 04:12:19.71 & $-$05:54:48.6 & 0.130 & GMOS+OSIRIS & 11.01 & 0.20 & 0.10 & 0.05 \\ 
  G08-5 & 08:54:18.74 & +06:46:20.5 & 0.132 & GMOS & 10.57 & 0.20 & 0.00 & 0.05 \\ 
  G10-1 & 10:21:42.47 & +12:45:18.8 & 0.144 & GMOS & 10.44 & 0.20 & 0.08 & 0.05 \\ 
  G20-2 & 20:04:42.92 & $-$06:46:57.9 & 0.141 & GMOS+OSIRIS & 10.61 & 0.20 & 0.14 & 0.05  \\ 
  SDSS 013527$-$1039 & 01:35:27.10 & $-$10:39:38.6 & 0.127 & OSIRIS & 10.83 & 0.20 & 0.11 & 0.05 \\ 
  SDSS 024921$-$0756 & 02:49:21.42 & $-$07:56:58.7 & 0.153 & OSIRIS & 10.48 & 0.20 & 0.73 & 0.05  \\ 
  SDSS 033244+0056 & 03:32:44.77 & +00:58:42.1 & 0.182 & GMOS & 10.86 & 0.20 & 0.00 & 0.05  \\ 
  SDSS 212912$-$0734 & 21:29:12.15 & $-$07:34:57.6 & 0.184 & OSIRIS & 10.85 & 0.20 & 0.00 & 0.05  \\ 
  SDSS 234657+0056 & 23:46:57.12 & +00:56:28.9 & 0.182 & GMOS+OSIRIS & 10.79 & 0.20 & 0.05 & 0.05   \\ 
  COSMOS 127977 & 09:59:37.961 & 02:18:02.16 & 1.62 & KMOS+OSIRIS & 10.77 & 0.20 & 0.00 & 0.10  \\ 
  UDS 78317 & 02:17:34.193 & -05:10:16.61 & 1.47 & KMOS+OSIRIS & 10.32 & 0.20 & 0.00 & 0.10  \\ 
   \hline
\end{tabular}\\
\begin{tabular}{lrrrrrrrr}
  \hline
Name & $r_d$ & $r_{flat}$ & $v_{flat}$ & $j_*$ & $\Delta j_*$ & $\tilde{j_*}$ & $\sigma$ & $\Delta \sigma$ \\ 
      &  [kpc] & [kpc] & [km s$^{-1}$] & [kpc km s$^{-1}$] & [kpc km s$^{-1}$] & [kpc km s$^{-1}$] & [km s$^{-1}$] & [km s$^{-1}$]\\
(1)  & (10) & (11) & (12) & (13) & (14) & (15) & (16) & (17)\\
  \hline
  C22-2 &  1.5 & 1.3 & 164 & 449 & 53 & 498 & 32 & 5 \\ 
  D13-5 & 1.3 & 0.4 & 171 & 466 & 51 & 437 & 46 & 5 \\ 
  G04-1 &  2.3 & 0.7 & 221 & 991 & 118 & 1025 & 50 & 5 \\ 
  G08-5 &  1.3 & 0.5 & 248 & 733 & 65 & 660 & 64 & 5 \\ 
  G10-1 & 3.2 & 1.1 & 118 & 670 & 105 & 756 & 52 & 3 \\ 
  G20-2  & 1.1 & 0.7 & 121 & 309 & 19 & 277 & 81 & 5 \\ 
  SDSS 013527$-$1039 &  1.8 & 1.0 & 118 & 404 & 53 & 429 & 41 & 5 \\ 
  SDSS 024921$-$0756 &  1.1 & 0.4 & 84 & 188 & 31 & 185 & 57 & 5 \\ 
  SDSS 033244+0056 & 1.5 & 0.7 & 239 & 804 & 59 & 695 & 59 & 5 \\ 
  SDSS 212912$-$0734 &  1.5 & 0.6 & 101 & 317 & 47 & 303 & 53 & 5 \\ 
  SDSS 234657+0056 &2.5 & 0.7 & 85 & 404 & 41 & 428 & -- & -- \\ 
  COSMOS 127977 &  4.8 & 3.9 & 236 & 2532 & 227 & 2279 & 57 & 11 \\ 
  UDS 78317 &  3.5 & 0.2 & 29 & 338 & 64 & 200 & 60 & 23 \\ 
   \hline
\end{tabular}\\
Columns: (1) galaxy identifier;  (2) right ascension (J2000);
 (3) declination (J2000);
 (4) redshift;
 (5) adopted data source (natural seeing GMOS or KMOS, adaptive optics OSIRIS, or combination of natural seeing and adaptive optics);
 (6) base 10 logarithm of stellar mass; (7) measurement uncertainty in $M_*$; (8) bulge-to-total ratio; (9) measurement uncertainty in $\beta$; (10) exponential disk scale length; (11) radius at which rotation curve becomes flat; (12) asymptotic velocity; (13) stellar specific AM; (14) measurement uncertainty in $j_*$; (15) approximate stellar specific AM using $j=krv$; (16) velocity dispersion; (17) measurement uncertainty in $\sigma$.
\end{table*}



\bibliographystyle{mnras}
\bibliography{references} 






\bsp    
\label{lastpage}
\end{document}